\def\draftversion{false} % true: single column, false: double column
\def\showall{true}  % Set to false to hide figures

\RequirePackage{ifthen}
\ifthenelse{\equal{\draftversion}{true}}{
  \documentclass[
  aps,
  prb,
  10pt,
  galley,
  amsmath,amssymb,
  superscriptaddress,
  nofootinbib,notitlepage,
  showpacs,
  preprintnumbers
  ]{revtex4-1}
\usepackage{showlabels}
}{
  \documentclass[
  aps,
  prb,
  10pt,
  twocolumn,
  amsmath,
  amssymb,
  superscriptaddress,
  showpacs,
  preprintnumbers
  ]{revtex4-2}
}

\usepackage{MSGstyle}
\usepackage{my_comment}
\usepackage{gensymb}
\usepackage{tabularray}
\usepackage{makecell}
\usepackage{orcidlink}
\usepackage{ulem}
\usepackage{cancel}

\newcommand{\vseff}{\vb{s}^{\rm eff}}
\newcommand{\vjeff}{\vb{J}^{\rm eff}}

\newcommand{\oseff}{\hat{\vb{s}}^{\rm eff}}

\newcommand{\jeff}{J_{\rm eff}}
\newcommand{\aj}{\mathcal{J}}

\begin{document}

\title{First-principles study of bulk stacking, $J_\textrm{eff}$ picture, magnetic Hamiltonian, $g$ factors, and structural distortions of $\alpha$-RuCl$_3$}
%\title{Bulk stacking, {\color{red}$J_\textrm{eff}$ picture revisited,} magnetic Hamiltonian, $g$ factors, and structural distortions of $\alpha$-RuCl$_3$: A comprehensive first-principles study}

\author{Seung-Ju Hong\,\orcidlink{0000-0002-9216-8657}}
\email{sjhong6230@snu.ac.kr}
\affiliation{Department of Physics and Astronomy, Seoul National University, Seoul 08826, Korea}
\affiliation{Center for Theoretical Physics, Seoul National University, Seoul 08826, Korea}
\author{Tae Yun Kim\,\orcidlink{0000-0001-8371-0654}}
\email{taeyun.kim@austin.utexas.edu}
\affiliation{Department of Physics and Astronomy, Seoul National University, Seoul 08826, Korea}
\affiliation{Center for Theoretical Physics, Seoul National University, Seoul 08826, Korea}
\affiliation{Center for Correlated Electron Systems, Institute for Basic Science, Seoul 08826, Korea}
\affiliation{Oden Institute for Computational Engineering and Sciences, The University of Texas at Austin, Austin, Texas 78712, USA}
\affiliation{Department of Physics, The University of Texas at Austin, Austin, Texas 78712, USA}
\author{Cheol-Hwan Park\,\orcidlink{0000-0003-1584-6896}}
\email{cheolhwan@snu.ac.kr}
\affiliation{Department of Physics and Astronomy, Seoul National University, Seoul 08826, Korea}
\affiliation{Center for Theoretical Physics, Seoul National University, Seoul 08826, Korea}
\affiliation{Center for Correlated Electron Systems, Institute for Basic Science, Seoul 08826, Korea}

\date{\today}

\begin{abstract}
$\alpha$-RuCl$_3$ is a candidate Kitaev material that exhibits zigzag antiferromagnetic ordering below 7 K. One contentious issue regarding this material is its bulk structure in the low-temperature phase. Recently, it has become generally accepted from experiments that the low- and high-temperature structures belong to the $R\bar{3}$ and $C2/m$ space groups, respectively. However, there was no theoretical study supporting the $R\bar{3}$-type structure as the low-temperature structure. In this study, we use constrained density functional theory to show that the $R\bar{3}$ structure is lower in energy than the $C2/m$ structure, in agreement with experimental observations.
Then, we show that the conduction band minimum states are almost of the $J_\textrm{eff}=1/2$ and $m_\textrm{eff}=-1/2$ character, if we set the angular momentum quantization axis to be parallel to the N\'eel vector; this is the first analysis of the $J_\textrm{eff}$ picture for $\alpha$-RuCl$_3$ from this perspective.
In addition, we compute the anisotropic magnetic exchange parameters and $g$ factors of monolayer $\alpha$-RuCl$_3$, thereby providing a comprehensive understanding of its magnetism. Our results demonstrate that both second-nearest-neighbor exchange interactions and magnetic moments not captured by the conventional atomic orbital projection method are necessary for accurate description of the magnetism in $\alpha$-RuCl$_3$. Moreover, the calculated $g$ factors are in fairly good agreement with experimental measurements, especially the small anisotropy between their in-plane and out-of-plane components. Finally, we examine the effects of structural distortions from a perfect RuCl$_6$ octahedron, already present in bulk $\alpha$-RuCl$_3$ without any external perturbation, on the magnetic properties. Our results indicate that relative twist of the upper and lower Cl triangles in the RuCl$_6$ octahedron significantly influences the magnetic anisotropy, suggesting a novel way of engineering the magnetism in this class of materials. % Our methods can be applied to a wide range of magnetic systems, making them generally useful for determining the magnetic structures and Hamiltonians of many materials.
% One of the widely used methods for the investigation of magnetic properties is the magnetic model Hamiltonian. It contains the exchange parameters, which determine the coupling between magnetic moments, and the $g$ factors, which determine the coupling to the external magnetic field. Firstly, from relaxation calculations with spin-orbit coupling and constrained moments, we found that the $R\bar{3}$ stacking has lower energy compared to $C2/m$ stacking, consistent with recent reports. Then, by using energy mapping, we obtained \textit{ab initio} magnetic exchange parameters for spin-orbit coupled system, $\alpha$-RuCl$_3$. By examining different models, we were able to deduce that inclusion of second nearest neighbor exchanges are needed. Furthermore, by canting the $\jeff${} moments, we computed the magnetic $g$ factor from the computation of total spin and orbital moments. From the results, we concluded that the consideration of itinerant moments is necessary for the computation of the $g$ factor. Also, by examining the effects of various structures, we found out that distortions beyond trigonal one is needed for qualitative description of magnetic properties and some exchange parameters are sensitive to the structural changes. Our methods can be applied to general magnetic systems and thus can be generally used to determine the magnetic structures and Hamiltonians of various materials.
\end{abstract}

\maketitle
\section{Introduction}

Recently, magnetism in two-dimensional van der Waals materials has been widely investigated both theoretically and experimentally~\cite{burch_magnetism_2018}. The first monolayer exfoliation of NiPS$_3$~\cite{kuo_exfoliation_2016} and the discovery of Ising-type antiferromagnetism in monolayer FePS$_3$~\cite{FePS3_AFM} opened a vast field, van der Waals magnetism. Monolayer magnetic materials are a good platform for investigating how magnetic properties vary under different conditions, as they have many tuning knobs, such as strain, gating, light, and proximity. Furthermore, monolayer magnetic materials can be used as building blocks for various heterostructures
\cite{burch_magnetism_2018}.

Another important subject in the field of magnetism is quantum spin liquid states~\cite{savary_quantum_2016, RevModPhys.89.025003}. The quantum spin liquids are a nontrivial state of magnetic systems with many-body quantum entanglement, where long-range order is suppressed because of strong quantum fluctuations.

The Kitaev honeycomb model is a spin-$1/2$ model defined on a two-dimensional honeycomb lattice~\cite{kitaev_anyons_2006}. The ground state of this model is known as the Kitaev spin liquid, which shows many physically interesting properties, such as fractionalized quasiparticles and Abelian/non-Abelian anyons~\cite{kitaev_anyons_2006}. Also, the quantum-entangled nature of the Kitaev spin liquid can be used in topological quantum computing~\cite{kitaev_anyons_2006}. 

There have been numerous works to realize the Kitaev spin liquid in real materials. Many of such attempts were based on the Jackeli-Khaliullin mechanism~\cite{jackeli_mott_2009}.
According to this mechanism, a honeycomb lattice with transition-metal complexes can host a magnetic state with strong spin-orbital correlation, namely the $\jeff${} state.
The occurrence of this exotic state requires a combination of strong spin-orbit coupling and localized states, which can be found in the $4d$ and $5d$ transition-metal compounds with five electrons in the $d$ orbitals~\cite{jeff_RuCl3, jeff_half_state}.
% arise from the $t_{2g}$ orbitals, when a strong spin-orbit coupling comparable to the octahedral crystal field is present.
In these materials, the four low-energy states ($\jeff=3/2$) are fully occupied, while the two high-energy states ($\jeff=1/2$) are half-filled, resulting in a nonvanishing effective spin-1/2 moment at each transition-metal-ligand octahedron.
Unlike the spin-only moments that appear in some $3d$ transition-metal antiferromagnets, such as NiPS$_3$ and MnPS$_3$~\cite{kimMagneticAnisotropyMagnetic2021}, the effective spin moments arising from the half-filled $\jeff=1/2$ states can interact in a highly anisotropic way. Jackeli and Khaliullin showed that when the octahedra are connected in the edge-sharing geometry and the direct hopping between $d$ orbitals is neglected, a Kitaev magnetism can emerge~\cite{jackeli_mott_2009}.

$\alpha$-RuCl$_3$ has been considered as a promising candidate that can realize the Jackeli-Khaliullin mechanism. However, direct exchange and further superexchange interactions stabilize a zigzag antiferromagnetic order below 7~K in bulk $\alpha$-RuCl$_3$~\cite{caoLowtemperatureCrystalMagnetic2016}, a feature that would be absent in a real Kitaev quantum spin liquid. There are experimental observations that a strong in-plane magnetic field can lead to a half-quantized thermal Hall effect due to the appearance of a Kitaev quantum spin liquid state~\cite{RuCl3_half_quantized}, but such observations depend strongly on the quality of samples and may not be directly related to Kitaev magnetism~\cite{xing_magnetothermal_2025}.

A crucial factor determining the magnetism of a material is its crystal structure. There have been debates on whether the low-temperature structure of bulk $\alpha$-RuCl$_3$ belongs to $C2/m$~\cite{caoLowtemperatureCrystalMagnetic2016} or $R\bar{3}$~\cite{parkEmergenceIsotropicKitaev2024} space group. Recent experimental works reported that the $R\bar{3}$ stacking is the low-temperature structure, and a transition to the $C2/m$ stacking occurs at about 150~K~\cite{parkEmergenceIsotropicKitaev2024, RuCl3_R3, RuCl3_R3_2, RuCl3_R3_3}.
In contrast, there has been no systematic theoretical study that addresses this issue.

In addition, many fundamental magnetic properties of $\alpha$-RuCl$_3$ are still under debate, both theoretically and experimentally.
One notable issue is the $g$ factor anisotropy.
As shown in Table~\ref{tab:previous_g}, some papers reported small anisotropies~\cite{majumder_anisotropic_2015,PhysRevB.96.161107,loidl_proximate_2021} in the measured $g$ factors, while some experimental~\cite{kubota_successive_2015} and theoretical~\cite{yadavKitaevExchangeFieldinduced2016a} studies reported large anisotropies in the $g$ factor. 
Another unsettled issue lies in the magnetic model Hamiltonian for bulk $\alpha$-RuCl$_3$. 
The exchange interaction parameters reported from previous studies~\cite{banerjeeProximateKitaevQuantum2016b,cookmeyerSpinwaveAnalysisLowtemperature2018,eichstaedtDerivingModelsKitaev2019a,houUnveilingMagneticInteractions2017a,kim_crystal_2016,laurellDynamicalThermalMagnetic2020a,ozelMagneticFielddependentLowenergy2019,ranSpinWaveExcitationsEvidencing2017,sahasrabudheHighfieldQuantumDisordered2020,searsFerromagneticKitaevInteraction2020,suzukiEffectiveModelStrong2018,wangTheoreticalInvestigationMagnetic2017a,winterChallengesDesignKitaev2016,winterBreakdownMagnonsStrongly2017a,wuFieldEvolutionMagnons2018,yadavKitaevExchangeFieldinduced2016a} are summarized in Table~\ref{tab:previous_exchange}.
Interestingly, the nearest-neighbor Kitaev exchange $K_1$ was reported to be negative (favoring ferromagnetic state) in 19 papers, but positive (favoring antiferromagnetic state) in four papers.
While recent studies tend to agree that $K_1<0$, significant differences remain in the values of the parameters.

In this work, we examined the crystal structure of bulk $\alpha$-RuCl$_3$ and the magnetic Hamiltonian of its monolayer using constrained density functional theory (DFT). We performed structural relaxation calculations assuming different structures ($C2/m$ or $R\bar{3}$, separately) to determine the more stable structure. 
%Our results show that regardless of the Hubbard $U$ parameter, the $R\bar{3}$ structure is lower in energy, in agreement with recent experimental results~\cite{parkEmergenceIsotropicKitaev2024, RuCl3_R3, RuCl3_R3_2, RuCl3_R3_3}.

We also show that the conduction band minimum states, when projected to the $d$ states of Ru atoms, are almost of the $J_\textrm{eff}=1/2$ and $m_\textrm{eff}=-1/2$ character, if we set the angular momentum quantization axis to be parallel to the N\'eel vector. Therefore, we can gain much insight into the spin-orbital entangled states by properly choosing the angular-momentum quantization axis.

Then, we delved into the magnetism of monolayer $\alpha$-RuCl$_3$. We computed the anisotropic magnetic exchange parameters with the energy mapping method.
% There are many studies on the exchange parameters of $\alpha$-RuCl$_3$~\cite{banerjeeProximateKitaevQuantum2016b,cookmeyerSpinwaveAnalysisLowtemperature2018,eichstaedtDerivingModelsKitaev2019a,houUnveilingMagneticInteractions2017a,kim_crystal_2016,laurellDynamicalThermalMagnetic2020a,ozelMagneticFielddependentLowenergy2019,ranSpinWaveExcitationsEvidencing2017,sahasrabudheHighfieldQuantumDisordered2020,searsFerromagneticKitaevInteraction2020,suzukiEffectiveModelStrong2018,wangTheoreticalInvestigationMagnetic2017a,winterChallengesDesignKitaev2016,winterBreakdownMagnonsStrongly2017a,wuFieldEvolutionMagnons2018,yadavKitaevExchangeFieldinduced2016a}, which are tabulated in Tab.~\ref{tab:previous_exchange}. The Kitaev exchange $K$ is negative (favoring ferromagnetic state) in 19 papers and positive (favoring antiferromagnetic state) in 4 papers.
Our investigation shows that an accurate description of the magnetic properties requires inclusion of the second-nearest-neighbor exchange interaction, which has been neglected in previous studies.
% Another unsettled issue in the magnetic properties of $\alpha$-RuCl$_3$ is the $g$ factor anisotropy. We summarize previous studies on the $g$ factors in Tab.~\ref{tab:previous_g}. Some papers reported small anisotropies in the $g$ factor~\cite{majumder_anisotropic_2015,PhysRevB.96.161107,loidl_proximate_2021}, while others reported large anisotropies~\cite{kubota_successive_2015,yadavKitaevExchangeFieldinduced2016a}.

Furthermore, we computed the $g$ factors of monolayer $\alpha$-RuCl$_3$ by canting the $\jeff$ moment through constrained DFT calculations.
% and computation of the induced spin/orbital magnetic moments.
% Especially, 
For an accurate evaluation of the orbital magnetic moments, we used the translationally equivariant Wannier interpolation method, which some of the authors recently developed~\cite{transl_inv}.
We show that the anisotropy in the $g$ factor is small
% (Tab.~\ref{table:moments})
and that the conventional atomic orbital projection method significantly underestimates the magnitude of the $g$ factor.

Lastly, we examined the effects of structural distortions from a perfect RuCl$_6$ octahedron, which are present in the actual $\alpha$-RuCl$_3$, on the magnetic properties of the monolayer. We considered two types of distortion on the RuCl$_6$ perfect octahedron: trigonal distortion and relative twist of the upper Cl triangle and the lower Cl triangle in the octahedron. Note that both are present in both the experimentally measured and computationally relaxed structures. We demonstrate that the latter (the relative twist), which has received no attention so far, has a greater impact on the magnetic properties of $\alpha$-RuCl$_3$, and thus cannot be neglected.% Our methods can be extended to various magnetic systems and thus have wide applicability.

\begin{table}[ht!]
\centering
\caption{The $g$ factor of bulk $\alpha$-RuCl$_3$ reported in different papers}
\begin{tblr}{
    cells = {c},
    hline{1,7} = {-}{0.08em},
    hline{2} = {-}{0.04em}
}
\makecell{Reference \\ (Publication Date)} & Method & $g_{XY}$ & $g_{Z}$ \\
\makecell{\citet{kubota_successive_2015} \\ (2015.03.23)} & \makecell{High-field\\magnetization} & 2.5 & 0.40 \\
\makecell{\citet{majumder_anisotropic_2015} \\ (2015.05.07)} & \makecell{High-temperature\\susceptibility} & 2.8 & 2.8 \\
\makecell{\citet{yadavKitaevExchangeFieldinduced2016a} \\ (2016.11.30)} & {Quantum chemistry\\calculation} & 2.51 & 1.09 \\
\makecell{\citet{PhysRevB.96.161107} \\ (2017.10.12)} & \makecell{X-ray absorption\\spectroscopy + \\multiplet calculation} & 2.27 & 2.05 \\
\makecell{\citet{loidl_proximate_2021} \\ (2021.08.27)} & \makecell{High-temperature\\susceptibility} & 2.9 & 2.4 \\
\end{tblr}
\label{tab:previous_g}
\end{table}

\section{Methods} \label{sec:methods}

% TYK comment
\begin{table*}[ht!]
\centering
\caption{Magnetic exchange parameters of bulk $\alpha$-RuCl$_3$ from various papers. ED, LSWT, INS, and SOC stand for exact diagonalization, linear spin wave theory, inelastic neutron scattering, and spin-orbit coupling, respectively. $P3_112$ and $C2/m$ denote the assumed space group. In all studies, $\oseff_i = \boldsymbol{\sigma}/2$ is assumed, where $\oseff_i$ is the effective spin operator at atom $i$ and $\boldsymbol{\sigma}$ is the Pauli matrix.
For the definition of the operator and exchange parameters ($J_1$, $K_1$, $\Gamma_1$, $\Gamma'_1$, and $J_3$), refer to Sec.~\ref{sec:methods}.}
\begin{tblr}{
  cells = {c},
  hline{1,25} = {-}{0.08em},
  hline{2} = {-}{0.04em}
}
\makecell{Reference \\ (Publication Date)} & Method & $J_1$ & $K_1$ & $\Gamma_1$ & $\Gamma'_1$ & $J_3$ \\
\makecell{\citet{banerjeeProximateKitaevQuantum2016b} \\ (2016.04.04)} & LSWT+INS fit  & $-$4.60 & +7.00 &          &           &     \\
\makecell{\citet{kim_crystal_2016} \\ (2016.04.22)} & (1) DFT+SOC+$t/U$, $P3_112$ & $-$3.50 & +4.60 & +6.42 & $-$0.04 & \\
& \makecell{(1)+Variable-cell relax \\ without magnetism and $U$} &  $-$2.76 & $-$3.55 & +7.08 & $-$0.54 & \\
& \makecell{(1)+Fixed-cell relax \\ without magnetism and $U$} &  $-$1.53 & $-$6.55 & +5.25 & $-$0.95 & \\
& \makecell{(1)+Variable-cell relax\\with zigzag AFM \\ and DFT+$U$} &  $-$0.97 & $-$8.21 & +4.16 & $-$0.93 & \\
\makecell{\citet{winterChallengesDesignKitaev2016} \\ (2016.06.27)} & DFT+ED, $P3_112$ & $-$5.50 & +7.60 & +8.40 & +0.20 & +2.30 \\ 
& DFT+ED, $C2/m$ & $-$1.67 & $-$6.67 & +6.60 & $-$0.87 & +2.80 \\ 
\makecell{\citet{yadavKitaevExchangeFieldinduced2016a} \\ (2016.11.30)} & Quantum chemistry &  +1.20 & $-$5.60 & $-$0.87 & & \\
\makecell{\citet{ranSpinWaveExcitationsEvidencing2017} \\ (2017.03.10)} & LSWT+INS fit &   & $-$6.80 & +9.50 & & \\
\makecell{\citet{houUnveilingMagneticInteractions2017a} \\ (2017.08.07)} & \makecell{DFT+mapping, $C2/m$,\\$U=3.0$ eV}  & $-$1.93 & $-$12.23 & +4.83 & & +1.60 \\
& \makecell{$U=3.5$ eV}  & $-$1.73 & $-$10.67 & +3.80 & & +1.27 \\
\makecell{\citet{wangTheoreticalInvestigationMagnetic2017a} \\ (2017.09.05)} & DFT+$t/U$, $P3_112$  & +0.10 & $-$5.50 & +7.60 & & +0.10 \\
& DFT+$t/U$, $C2/m$ &  $-$0.30 & $-$10.9 & +6.10 & & +0.03 \\
\makecell{\citet{winterBreakdownMagnonsStrongly2017a} \\ (2017.10.27)} & ED+INS fit &  $-$0.50 & $-$5.00 & +2.50 & & +0.50 \\
\makecell{\citet{suzukiEffectiveModelStrong2018} \\ (2018.04.25)} & ED+specific heat fit &  $-$1.53 & $-$24.41 & +5.25 & $-$0.95 &  \\
\makecell{\citet{cookmeyerSpinwaveAnalysisLowtemperature2018} \\ (2018.08.27)} & Thermal Hall fit  & $-$0.50 & $-$5.00 & +2.50 &  & +0.11 \\
\makecell{\citet{wuFieldEvolutionMagnons2018} \\ (2018.09.26)} & LSWT, THz fit  & $-$0.35 & $-$2.8 & +2.4 &  & +0.34 \\
\makecell{\citet{ozelMagneticFielddependentLowenergy2019} \\ (2019.08.02)} & \makecell{LSWT, THz fit,\\$K>0$} &  $-$0.95 & +1.15 & +2.92 &  &  \\
 & \makecell{LSWT, THz fit,\\$K<0$} &  &  & +0.46 & $-$3.50 & +2.35 &  &  \\
\makecell{\citet{eichstaedtDerivingModelsKitaev2019a} \\ (2019.08.06)} & DFT+Wannier+$t/U$ &  $-$1.40 & $-$14.3 & +9.80 & $-$2.23 & +0.97 \\
\makecell{\citet{laurellDynamicalThermalMagnetic2020a} \\ (2020.01.10)} & \makecell{ED+specific heat fit}  & $-$1.30 & $-$15.1 & +10.1 & $-$0.12 & +0.90 \\
\makecell{\citet{searsFerromagneticKitaevInteraction2020} \\ (2020.04.13)} & Magnetization fit  & $-$2.70 & $-$10.0 & +10.6 & $-$0.90 &  \\
\makecell{\citet{sahasrabudheHighfieldQuantumDisordered2020} \\ (2020.04.24)} & ED+Raman fit  & $-$0.75 & $-$10.0 & +3.75 &  & +0.75 \\
\end{tblr}
\label{tab:previous_exchange}
\end{table*}

We examine three types of model Hamiltonians: (1) the conventional Kitaev-Heisenberg (KH) model, (2) the spin-ice (SI) model proposed in Ref.~\onlinecite{maksimov_rethinking_2020}, and (3) our anisotropic $J_1$-$J_2$-$J_3$ (A-$J_1J_2J_3$) model. The Hamiltonians for the KH and SI models are shown below:
\begin{align}
    \hat{H} &= \sum_{\langle ij \rangle} \oseff_i \cdot \aj_{1,ij}\, \oseff_j  + J_3\sum_{\langle\langle\langle ij \rangle\rangle\rangle} \oseff_i \cdot \oseff_j \quad
    \label{eq:hamiltonian1}
\end{align}
where  $\oseff_i$ is the {\it effective spin} operator at atom $i$, and $\langle ij \rangle$ and $\langle\langle\langle ij \rangle\rangle\rangle$ denote the first- and third-nearest-neighbor pairs, respectively.
% , and $\tilde{\vb{J}}^i_{\rm eff}$ is the spin-1/2 operators acting on the two $J_\text{eff} = 1/2$ states of atom $i$.

\begin{figure}[t]
    \centering
    \includegraphics[width=0.5\textwidth]{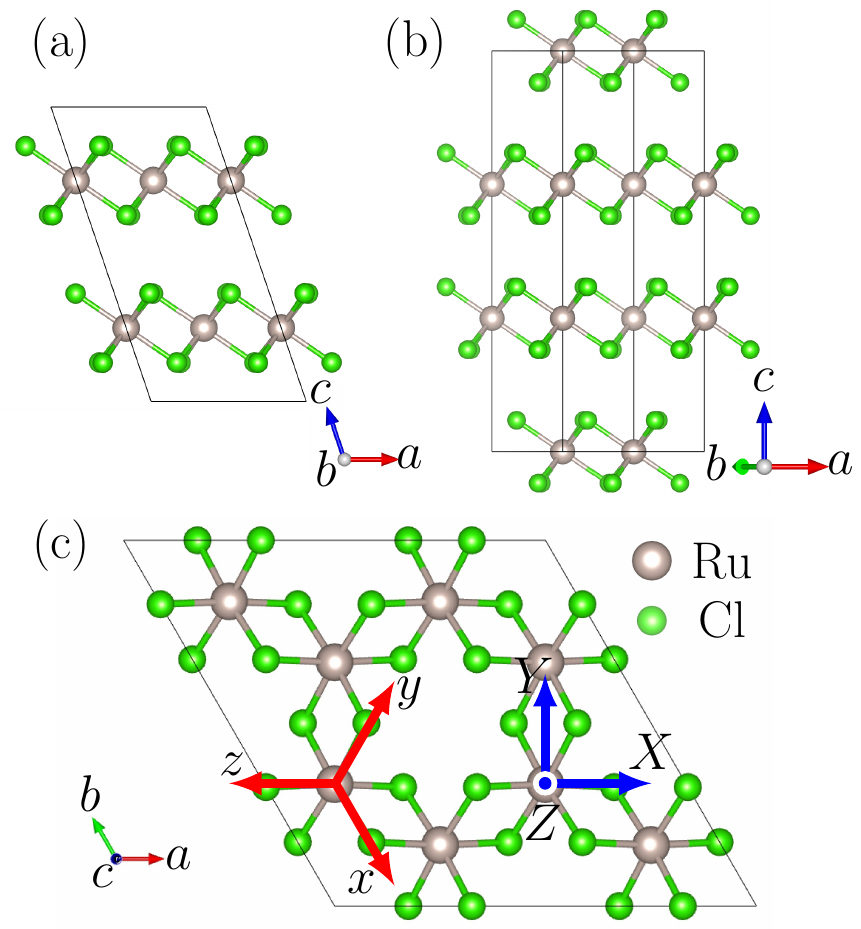}
    \caption{
    Crystal structure of bulk $\alpha$-RuCl$_3$ with (a) $C2/m$ stacking and (b) $R\bar{3}$ stacking.
    (c) Crystal structure of monolayer $\alpha$-RuCl$_3$. The $xyz$ axes correspond to the local octahedron axes, while the $XYZ$ axes are the global Cartesian axes.
    The $abc$ axes represent the crystallographic axes.
    }
    \label{fig:crystal_structure}
\end{figure}

Typically, the nearest-neighbor exchange tensor of the KH model is represented in the local octahedron $xyz$ coordinate system [Fig.~\ref{fig:crystal_structure}(c)].
In this convention, the exchange tensor of the KH model for nearest-neighbor Ru-Ru pairs along the $Y$ direction [Fig.~\ref{fig:crystal_structure}(c)] is written as
\begin{equation}
    \aj^{\text{KH}}_{1,ij} = 
    \begin{pmatrix}
        J_1 & \Gamma_1 & \Gamma'_1 \\
        \Gamma_1 & J_1 & \Gamma'_1 \\
        \Gamma'_1 & \Gamma'_1 & J_1 + K_1
    \end{pmatrix}\,,
    \label{eq:octahedron}
\end{equation}
where $J_1$ is the isotropic exchange, $K_1$ is the Kitaev exchange, $\Gamma_1$ and $\Gamma'_1$ are symmetric off-diagonal exchanges. Note that for ideal Jackeli-Khaliullin mechanism, only the $K_1$ term survives.

In the global Cartesian $XYZ$ coordinate system, $\aj^{\text{KH}}_{1,ij}$ is parametrized differently, i.e.,
\begin{equation}
    \aj^{\text{KH}}_{1,ij} = 
    \begin{pmatrix}
        J_{1,XX} & 0 & J_{1,ZX} \\
        0 & J_{1,YY} & 0 \\
        J_{1,ZX} & 0 & J_{1,ZZ}
    \end{pmatrix}\,.
    \label{eq:crystallographic}
\end{equation}
We note that this representation of $\aj^{\text{KH}}_{1,ij}$ is the most general form allowed by the $C_2$ symmetry associated with the $Y$ direction Ru-Ru pair.
The exchange tensors for the other two nearest-neighbor pairs can be obtained by transforming the $\aj^{\text{KH}}_{1,ij}$ in Eq.~(\ref{eq:crystallographic}),
as they are connected by the $C_3$ rotations with respect to the $Z$ axis;
the other nearest-neighbor exchange tensors are obtained with $R^{-1} \aj^{\text{KH}}_{1,ij} R$
and $R \aj^{\text{KH}}_{1,ij} R^{-1}$, where
\begin{equation}
    \label{eq:rot_120}
    R
    = \mqty(-1/2 & -\sqrt{3}/2 & 0 \\ \sqrt{3}/2 & -1/2 & 0 \\ 0 & 0 & 1),
\end{equation}
a 120$^\circ$ rotation matrix in the $XYZ$ coordinate system.

In the SI model, the same $Y$-bond nearest-neighbor exchange tensor in the global Cartesian $XYZ$ axis is
\begin{equation}
    \aj^{\text{SI}}_{1,ij} = 
    \begin{pmatrix}
        J_{1,XX} & 0 & J_{1,ZX} \\
        0 & J_{1,XX} & 0 \\
        J_{1,ZX} & 0 & 0
    \end{pmatrix},
    \label{eq:model2}
\end{equation}
which has two independent exchange parameters, fewer than four parameters in the KH model.

Finally, in the A-$J_1J_2J_3$ model, we include anisotropic second-nearest-neighbor exchange interactions. We also consider the anisotropy in third-nearest-neighbor exchange interactions. The Hamiltonian for this model is
\begin{align}
    \hat{H} 
    &= \sum_{\langle ij 
    \rangle} \oseff_i \cdot \aj^{\text{A-}J_1J_2J_3}_{1,ij}\, \oseff_j + \sum_{\langle\langle ij \rangle\rangle} \oseff_i \cdot \aj^{\text{A-}J_1J_2J_3}_{2,ij}\, \oseff_j\nnnl 
    &+\sum_{\langle\langle\langle ij \rangle\rangle\rangle} \oseff_i \cdot \aj^{\text{A-}J_1J_2J_3}_{3,ij}\, \oseff_j\,,
    \label{eq:hamiltonian2}
\end{align}
where $\langle ij \rangle$, $\langle \langle ij \rangle\rangle$, and $\langle\langle\langle ij \rangle\rangle\rangle$ denote the first-, second-, and third-nearest-neighbor effective moment pairs, respectively.
In this model, all exchange tensors are of the form of Eq.~\eqref{eq:octahedron} for the local octahedron axes ($xyz$) or of Eq.~\eqref{eq:crystallographic} for the global Cartesian axes ($XYZ$).

We fit the three model Hamiltonians to the total energies of monolayer $\alpha$-RuCl$_3$ obtained from our \textit{ab initio} DFT calculations, extracting the exchange parameters. For the fitting, we used
% the following approximation
% for the expectation value of the A-$J_1J_2J_3$ Hamiltonian
% that is implicitly assumed in
the total energy mapping method~\cite{kimMagneticAnisotropyMagnetic2021,houUnveilingMagneticInteractions2017a, 2011mapping, 2013mapping, 2020mapping}:
\begin{widetext}
\begin{align}
    % \langle \Psi | \hat{H} | \Psi \rangle
    E_{\rm cDFT}
    % &= \sum_{\langle ij \rangle}
       % \langle \Psi | \oseff_i \cdot \aj^{\text{A-}J_1J_2J_3}_{1,ij} \oseff_j | \Psi \rangle 
    % + \sum_{\langle\langle ij \rangle\rangle}
       % \langle \Psi | \oseff_i \cdot \aj^{\text{A-}J_1J_2J_3}_{2,ij} \oseff_j | \Psi \rangle 
    % +\sum_{\langle\langle\langle ij \rangle\rangle\rangle} 
       % \langle \Psi | \oseff_i \cdot \aj^{\text{A-}J_1J_2J_3}_{3,ij} \oseff_j | \Psi \rangle \nnnl 
    &\approx
    \sum_{\langle ij \rangle}
        \vseff_i \cdot \aj^{\text{A-}J_1J_2J_3}_{1,ij} \, \vseff_j 
    + \sum_{\langle\langle ij \rangle\rangle}
        \vseff_i \cdot \aj^{\text{A-}J_1J_2J_3}_{2,ij} \, \vseff_j 
    +\sum_{\langle\langle\langle ij \rangle\rangle\rangle} 
        \vseff_i \cdot \aj^{\text{A-}J_1J_2J_3}_{3,ij} \, \vseff_j\,,
    \label{eq:model_approx}
\end{align}
\end{widetext}
where $\vseff_i = \langle \Psi | \oseff_i | \Psi \rangle$, representing the expectation value of the effective spin operator,
$\ket{\Psi}$ is the ground-state wavefunction, a single Slater determinant of Kohn-Sham (KS) wavefunctions, obtained from the constrained DFT (cDFT) calculation,
and $E_{\rm cDFT}$ is the total energy minus a constant that does not depend on $\vseff_i$'s.
% This approximation is implicitly assumed for the total energy mapping method~\cite{kimMagneticAnisotropyMagnetic2021,houUnveilingMagneticInteractions2017a, 2011mapping, 2013mapping, 2020mapping}.

To ensure the convergence of the DFT calculations to the targeted $\jeff${} states, we adopt the constrained DFT. The method for constraining the direction of the effective spin-orbital angular momentum is analogous to the one described in Ref.~\onlinecite{kimMagneticAnisotropyMagnetic2021}.

We define the Hubbard occupation matrix for atom $i$, $\rho_{i,mn\sigma\sigma'}$:
\begin{equation}
    \rho_{i,mn\sigma\sigma'}
    = \sum_{v\vb{k}}^{\rm occ} \braket{\phi_{i,m\sigma}}{\psi_{v\vb{k}}} \braket{\psi_{v\vb{k}}}{\phi_{i,n\sigma'}}\,,
\end{equation}
where $\ket{\psi_{v\vb{k}}}$ is the KS wave function in the valence band with band index $v$ and Bloch wavevector $\vb{k}$ and $\ket{\phi_{i,m\sigma}}$ is one of the 10 atomic spin-$d$ orbitals at atom $i$ in the home unit cell; $m$ is the orbital index $(d_{z^2}, d_{x^2-y^2}, d_{yz}, d_{xz}, d_{xy})$, and $\sigma$ is the spin index $(\uparrow, \downarrow)$.
Note that these $d$ orbitals are defined with respect to the local octahedron axis ($xyz$).

% Then, we evaluate $\vseff_i$ by using
% \begin{align}
%     \vseff_i
%     = \frac{\sum_{mn\sigma\sigma'} \left(\rho_{i,mn\sigma\sigma'} \mjeff_{i,nm\sigma'\sigma} \right)}{2\left|\sum_{mn\sigma\sigma'} \left(\rho_{i,mn\sigma\sigma'} \mjeff_{i,nm\sigma'\sigma} \right)\right|}
% \end{align}
% where $\mjeff_{i,nm\sigma'\sigma} = \langle \phi_{i,n\sigma'} |\ojeff_i| \phi_{i,m\sigma} \rangle$ is the matrix representation of the effective angular momentum operator $\ojeff_i = \oleff_i + \os_i$ in the spin-$d$ orbital basis. These operators and their matrix representations are defined in the following way.
In this local spin-orbital subspace, the matrix representation of the atomic orbital angular momentum operators $\hat{\bf L}_i = (\hat{\bf r} - {\bf r}_i) \times \hat{\bf p}$, where ${\bf r}_i$ is the position of the atom $i$,
% with respect to the center of atom $i$, $\hat{\vb{L}}_i = (\hat{\vb{r}} - \vb{r}_i) \times \hat{\vb{p}}$,
can be written as $\langle \phi_{i,m\sigma} |\hat{{\bf L}}_{i}| \phi_{i,n\sigma'} \rangle = {\bf L}_{mn} \delta_{\sigma\sigma'}$
{($\hat{{\bf L}}_{i} = \sum_{mn\sigma\sigma'} \ket{\phi_{i,m\sigma}} {\bf L}_{mn} \delta_{\sigma\sigma'} \bra{\phi_{i,n\sigma'}}$)},
where ${\bf L} = (L_x, L_y, L_z)$ and
\begin{align}
    L_{x} &=
        \left(
        \begin{array}{cc|ccc}
            0 & 0 & \sqrt{3}i & 0 & 0 \\
            0 & 0 & i & 0 & 0 \\
            \hline
            -\sqrt{3}i & -i & 0 & 0 & 0 \\
            0 & 0 & 0 & 0 & i \\
            0 & 0 & 0 & -i & 0
        \end{array}
        \right), \\
    L_{y} &=
        \left(
        \begin{array}{cc|ccc}
            0 & 0 & 0 & -\sqrt{3}i & 0 \\
            0 & 0 & 0 & i & 0 \\
            \hline
            0 & 0 & 0 & 0 & -i \\
            \sqrt{3}i & -i & 0 & 0 & 0 \\
            0 & 0 & i & 0 & 0
        \end{array}
        \right), \\
    L_{z} &=
        \left(
        \begin{array}{cc|ccc}
            0 & 0 & 0 & 0 & 0 \\
            0 & 0 & 0 & 0 & -2i \\
            \hline
            0 & 0 & 0 & i & 0 \\
            0 & 0 & -i & 0 & 0 \\
            0 & 2i & 0 & 0 & 0
        \end{array}
        \right).
    % L_{x} &= \mqty(0 & 0 & \sqrt{3}i & 0 & 0 \\
    %           0 & 0 & 0 & 0 & i \\
    %           -\sqrt{3}i & 0 & 0 & -i & 0 \\
    %           0 & 0 & i & 0 & 0 \\
    %           0 & -i & 0 & 0 & 0) \\
    % L_{y} &= \mqty(0 & -\sqrt{3}i & 0 & 0 & 0 \\
    %           \sqrt{3}i & 0 & 0 & -i & 0 \\
    %           0 & 0 & 0 & 0 & -i \\
    %           0 & i & 0 & 0 & 0 \\
    %           0 & 0 & i & 0 & 0) \\
    % L_{z} &= \mqty(0 & 0 & 0 & 0 & 0 \\
    %           0 & 0 & -i & 0 & 0 \\
    %           0 & i & 0 & 0 & 0 \\
    %           0 & 0 & 0 & 0 & -2i \\
    %           0 & 0 & 0 & 2i & 0).
\end{align}
The expectation value of the (atomic) orbital angular momentum given by
\begin{equation}
        {\bf L}_{i} = \sum_{mn\sigma\sigma'} \rho_{i,mn\sigma\sigma'} {\bf L}_{nm} \delta_{\sigma'\sigma}.
\end{equation}
% \begin{align}
%     \langle \phi_{i,n\sigma'} |\hat{L}_{x,i}| \phi_{i,m\sigma} \rangle &= 
%     \mqty(0 & 0 & \sqrt{3}i & 0 & 0 \\
%           0 & 0 & 0 & 0 & i \\
%           -\sqrt{3}i & 0 & 0 & -i & 0 \\
%           0 & 0 & i & 0 & 0 \\
%           0 & -i & 0 & 0 & 0) \otimes
%     \mqty(1 & 0 \\ 0 & 1) \\
%     \langle \phi_{i,n\sigma'} |\hat{L}_{y,i}| \phi_{i,m\sigma} \rangle &= 
%     \mqty(0 & -\sqrt{3}i & 0 & 0 & 0 \\
%           \sqrt{3}i & 0 & 0 & -i & 0 \\
%           0 & 0 & 0 & 0 & -i \\
%           0 & i & 0 & 0 & 0 \\
%           0 & 0 & i & 0 & 0) \otimes
%     \mqty(1 & 0 \\ 0 & 1) \\
%     \langle \phi_{i,n\sigma'} |\hat{L}_{z,i}| \phi_{i,m\sigma} \rangle &= 
%     \mqty(0 & 0 & 0 & 0 & 0 \\
%           0 & 0 & -i & 0 & 0 \\
%           0 & i & 0 & 0 & 0 \\
%           0 & 0 & 0 & 0 & -2i \\
%           0 & 0 & 0 & 2i & 0) \otimes
%     \mqty(1 & 0 \\ 0 & 1) \, ,
% \end{align}
% where the $5 \times 5$ matrices act on the orbital index $m, n$ and the $2 \times 2$ matrices act on the spin index $\sigma, \sigma'$.
Except for the overall sign the (bottom right) $t_{2g}$ blocks  of ${\bf L}$ are equal to the matrix representation of the angular momentum operators in the $L=1$ real spherical harmonics basis $(p_x, p_y, p_z)$.
This observation leads to the following matrix representation of the effective orbital angular momentum ${\bf L}^{\rm eff} = (L_x^{\rm eff}, L_y^{\rm eff}, L_z^{\rm eff})$:
\begin{align}
    L_{x}^{\rm eff} &=
        \left(
        \begin{array}{cc|ccc}
            0 & 0 & 0 & 0 & 0 \\
            0 & 0 & 0 & 0 & 0 \\
            \hline
            0 & 0 & 0 & 0 & 0 \\
            0 & 0 & 0 & 0 & -i \\
            0 & 0 & 0 & i & 0
        \end{array}
        \right), \\
    L_{y}^{\rm eff} &=
        \left(
        \begin{array}{cc|ccc}
            0 & 0 & 0 & 0 & 0 \\
            0 & 0 & 0 & 0 & 0 \\
            \hline
            0 & 0 & 0 & 0 & i \\
            0 & 0 & 0 & 0 & 0 \\
            0 & 0 & -i & 0 & 0
        \end{array}
        \right), \\
    L_{z}^{\rm eff} &=
        \left(
        \begin{array}{cc|ccc}
            0 & 0 & 0 & 0 & 0 \\
            0 & 0 & 0 & 0 & 0 \\
            \hline
            0 & 0 & 0 & -i & 0 \\
            0 & 0 & i & 0 & 0 \\
            0 & 0 & 0 & 0 & 0
        \end{array}
        \right).
\end{align}

The matrix representation of the atomic spin operator is given by $\langle \phi_{i,m\sigma} |\hat{{\bf S}}_{i}| \phi_{i,n\sigma'} \rangle = \frac{\boldsymbol{\sigma}_{\sigma\sigma'}}{2} \delta_{mn}$ {($\hat{{\bf S}}_{i} = \sum_{mn\sigma\sigma'} \ket{\phi_{i,m\sigma}} \frac{\boldsymbol{\sigma}_{\sigma\sigma'}}{2} \delta_{mn} \bra{\phi_{i,n\sigma'}}$)},
where $\boldsymbol{\sigma} = (\sigma_x, \sigma_y, \sigma_z)$ and
\begin{equation}
    \sigma_x = \mqty(0 & 1 \\ 1 & 0)\,,
    \sigma_y = \mqty(0 & -i \\ i & 0)\,,
    \sigma_z = \mqty(1 & 0 \\ 0 & -1),
\end{equation}
i.e., the Pauli matrices.
The expectation value of the (atomic) spin can be computed with
\begin{equation}
        {\bf S}_{i} = \sum_{mn\sigma\sigma'} \rho_{i,mn\sigma\sigma'} \frac{\boldsymbol{\sigma}_{\sigma\sigma'}}{2} \delta_{mn}.
\end{equation}

% The $\os_i$ operator is the projection of the spin operator $\hat{\vb{S}}$ to the spin-$d$ orbital basis of atom $i$. The matrix representations of these operators in the spin-$d$ orbital basis are
% \begin{equation}
%     \langle \phi_{i,n\sigma'} |\hat{\vb{S}}_{i}| \phi_{i,m\sigma} \rangle = 
%     \mqty(1 & 0 & 0 & 0 & 0 \\
%           0 & 1 & 0 & 0 & 0 \\
%           0 & 0 & 1 & 0 & 0 \\
%           0 & 0 & 0 & 1 & 0 \\
%           0 & 0 & 0 & 0 & 1) \otimes
%     \frac{\boldsymbol{\sigma}}{2}\, ,
% \end{equation}
% where $\boldsymbol{\sigma}$ are the Pauli matrices.

The matrix representation of the effective angular momentum is defined as ${\bf J}^{\rm eff}_{mn\sigma\sigma'} = \langle \phi_{i,m\sigma} |\hat{{\bf J}}^{\rm eff}_{i}| \phi_{i,n\sigma'} \rangle = {\bf L}^{\rm eff}_{mn} \delta_{\sigma\sigma'} + \frac{\boldsymbol{\sigma}_{\sigma\sigma'}}{2} \delta_{mn}$
 {($\hat{{\bf J}}_{i}^{\rm eff} = \sum_{mn\sigma\sigma'} \ket{\phi_{i,m\sigma}} {\bf J}_{mn\sigma\sigma'}^{\rm eff} \bra{\phi_{i,n\sigma'}}$)}.
% which can be formally written as ${\bf J}^{\rm eff} = {\bf L}^{\rm eff} + {\bf S}$.
Then, the expectation value of the effective (atomic) total angular momentum can be evaluated with
\begin{align}
    {\bf J}^{\rm eff}_{i} = \sum_{mn\sigma\sigma'} \rho_{i,mn\sigma\sigma'} {\bf J}^{\rm eff}_{nm\sigma'\sigma}.
\end{align}
One subtle issue with this expression for the effective spin degrees of freedom is that the magnitude of ${\bf J}^{\rm eff}_{i}$ defined in this way can deviate from $1/2$.
If ${\bf J}^{\rm eff}_{i}$ is equated with ${\bf s}^{\rm eff}_{i}$ [Eq.~(\ref{eq:model_approx})], the normalization of the exchange interaction parameters to be fitted in the next step can be affected.
We resolve this issue by defining $\vseff_i$ as
\begin{align} \label{eq:def_s_eff}
    \vseff_i
    = \frac{{\bf J}^{\rm eff}_{i}}{2|{\bf J}^{\rm eff}_{i}|},
\end{align}
ensuring that $\vseff_i$ is normalized with $|\vseff_i| = 1/2$.
Apparently, $\vseff_i/|\vseff_i| = \vjeff_i/|\vjeff_i|$.
% As in Tab.~\ref{tab:previous_exchange}, the magnitude convention of the calculated exchange interactions is fixed by renormalizing the magnitude of the effective spin moment, i.e., $|\vseff_i|$, to $1/2$. 

% where $\mjeff_{i,nm\sigma'\sigma} = \langle \phi_{i,n\sigma'} |\ojeff_i| \phi_{i,m\sigma} \rangle$ is the matrix representation of the effective angular momentum operator $\ojeff_i = \oleff_i + \os_i$ in the spin-$d$ orbital basis. These operators and their matrix representations are defined in the following way.
% Note that we must distinguish $\oseff_i$ from $\os_i$.
% Note that we distinguish operators with tilde and atomic index $i$ from operators with hat and atomic index $i$. The former acts on the two $J_\text{eff}=1/2$ states of atom $i$, while the latter acts on the 10 $d$-orbitals. That is, $\vseff = \hat{P}_i^{1/2} \ojeff \hat{P}_i^{1/2}$ where $\hat{P}_i^{1/2}$ the projection operator to the $J_{\rm eff}=1/2$ states of atom $i$.

We now define the constraining energy as
\begin{equation}
\label{eq:Econstr}
    E_\text{constr} = \Lambda \sum_i \left ( 1-\frac{\vseff_i}{|\vseff_i|}\cdot \mathbf{e}_i^{[\xi]} \right ),
\end{equation}
where the sum is taken over the magnetic atoms, and $\vb{e}_i^{[\xi]}$ is the constraining direction for $\vseff_i$.
The additional contribution to the KS self-consistent potential is given by $\sum_i \sum_{mn\sigma\sigma'} \ket{\phi_{i,m\sigma}} V^{\text{constr}}_{i,mn\sigma\sigma'} \bra{\phi_{i,n\sigma'}}$, where
\begin{align}
\label{eq:Vconstr}
\begin{split}
    V^{\text{constr}}_{i,mn\sigma\sigma'} 
    &= \frac{\partial E_\text{constr}}{\partial \rho_{i,nm\sigma'\sigma}} \\ 
    &= \frac{\partial \vjeff_i}{\partial \rho_{i,nm\sigma'\sigma}} \cdot \frac{\partial E_\text{constr}}{\partial \vjeff_i} \\
    &= \frac{\Lambda \vjeff_{mn\sigma\sigma'}}{|\vjeff_i|} \cdot\left[ \frac{\vseff_i}{|\vseff_i |}\left ( \frac{\vseff_i}{|\vseff_i|} \cdot \mathbf{e}_i^{[\xi]} \right ) - \mathbf{e}_i^{[\xi]} \right] \, .
\end{split}
\end{align}
We note that the above contribution was incorporated into the calculation of forces and stress needed for relaxation.
In other words, the direction of $\vseff_i$'s was kept the same during the structural relaxation calculations.
% In the calculation of the force and stress needed for relaxation, the constraining Hubbard potential [Eq.~\eqref{eq:Vconstr}] has to be added to the total Hubbard potential.
% $\vseff_i/|\vseff_i|$ converges to $\vb{e}_i$ with a sufficiently large $\Lambda$.

With the above contribution, the self-consistent field cycle converges to a desired effective spin configuration $[\xi]$. We define $| \psi_{v\vb{k}}^{[\xi]}\rangle$ as the Kohn-Sham wave function of the target configuration in the valence. We also define the Hubbard occupation matrix $\rho_{i,mn\sigma\sigma'}^{[\xi]}$ and the expectation value of the atomic orbital and spin moments, ${\bf L}_i^{[\xi]}$ and ${\bf S}_i^{[\xi]}$ respectively, for this system.
\begin{equation}
    \rho_{i,mn\sigma\sigma'}^{[\xi]}
    = \sum_{v\vb{k}}^{\rm occ} \langle{\phi_{i,m\sigma}} | {\psi_{v\vb{k}}^{[\xi]}} \rangle \langle {\psi_{v\vb{k}}^{[\xi]}} | {\phi_{i,n\sigma'}} \rangle \,,
\end{equation}
\begin{equation}
        {\bf L}_{i}^{[\xi]} = \sum_{mn\sigma} \rho_{i,mn\sigma\sigma}^{[\xi]} {\bf L}_{nm},
\end{equation}
\begin{equation}
        {\bf S}_{i}^{[\xi]} = \sum_{m\sigma\sigma'} \rho_{i,mm\sigma\sigma'}^{[\xi]} {\bf S}_{\sigma'\sigma}.
\end{equation}

\begin{figure}[t]
    \centering
    \includegraphics[width=0.5\textwidth]{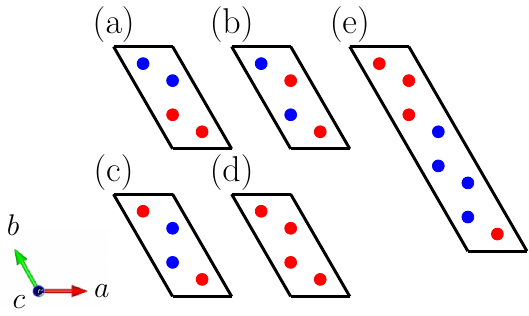}
    \caption{Magnetic configurations used for the computation of exchange parameters. Red and blue discs denote up and down effective spins, respectively, with respect to a given collinear direction. (a)-(d) correspond to zigzag, N\'eel, stripy, and ferromagnetic configurations of the honeycomb lattice, respectively.}
    \label{fig:configurations}
\end{figure}

With this formalism, the total energies of monolayer $\alpha$-RuCl$_3$ are computed with various magnetic configurations (Fig.~\ref{fig:configurations}) and $\vseff_i$ collinear directions $\vb{e}^{[\xi]} = (\sin\theta \cos\phi, \sin\theta \sin\phi, \cos\theta)^{\mathsf{T}}$. These directions denote the effective spin direction of the red atoms in Fig.~\ref{fig:configurations}. We chose $\theta$ from $0\degree$ to $165\degree$ with the $15\degree$ spacing for the $\phi=0$ ($XZ$) plane, and for the $\theta=90\degree$ ($XY$) plane, we chose the same sampling as $\theta$. Then, the model parameters in the Hamiltonians shown in Eqs.~\eqref{eq:hamiltonian1} and~\eqref{eq:hamiltonian2} are fitted to the computed total energies.
%\begin{figure}[t]
%    \centering
%    \includegraphics[width=0.4\textwidth]{configurations.pdf}
%    \caption{Magnetic configurations used for the computation of exchange parameters. Red and blue circles denote up and down spins, respectively.}
%    \label{fig:configurations}
%\end{figure}
Figures~\ref{fig:configurations}(a)-~\ref{fig:configurations}(d) show zigzag, N\'eel, stripy, and ferromagnetic configurations of the honeycomb lattice, respectively. However, for the A-$J_1J_2J_3$ model, these configurations are not enough to resolve the third-nearest-neighbor anisotropic tensor $\aj_{3,ij}$. Thus, we include the configuration in Fig.~\ref{fig:configurations}(e). These configurations are analogous to those used in Ref.~\onlinecite{kimMagneticAnisotropyMagnetic2021}.

We note that the definition of $\mathbf{s}_i^\text{eff}$ [Eq.~\eqref{eq:def_s_eff}] is based on the implicit assumption that the magnitude of the $\mathbf{J}_\text{eff}$ moments is a constant regardless of their directions.

In Ref.~\onlinecite{kimMagneticAnisotropyMagnetic2021}, the magnetic dipole energy was considered in the spin Hamiltonian, and its contribution to the magnetic anisotropy was of the order of $\mu$eV's. However, in our work, the anisotropy arising from the exchange interaction is of the order of meV's.
%Also, the magnitude of the atomic spin moment is smaller than in Ref.~\onlinecite{kimMagneticAnisotropyMagnetic2021} because $|\vseff_i|$ is 1/2, smaller than 5/2, 2, and 1 used for MnPS$_3$, FePS$_3$, and NiPS$_3$, respectively, in Ref.~\onlinecite{kimMagneticAnisotropyMagnetic2021}.
Thus, we can safely ignore this contribution.

\begin{figure}[t]
    \centering
    \includegraphics[width=0.5\textwidth]{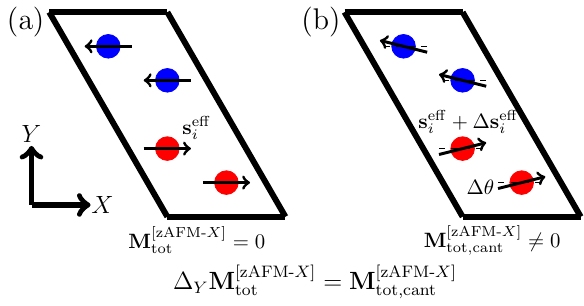}
    \caption{Schematic of the canting method for obtaining the $g$ factors. (a) Before canting, the magnetic moments are aligned along the $X$ direction, and the total magnetic moment ($\vb{M}_{\rm tot}^{[\text{zAFM-}X]}$) is zero. (b) After canting the moments slightly to the $Y$ direction (exaggerated in the figure), the $Y$ component of the total magnetic moment ($\vb{M}_{\rm tot, cant}^{[\text{zAFM-}X]}$) is proportional to the canting angle $\Delta \theta$. All structures are zigzag antiferromagnetic.}
    \label{fig:canting}
\end{figure}

For the computation of the $g$ factor of a magnetic moment in the monolayer, we tilted the direction of a $\jeff$ moment in the zigzag antiferromagnetic system 1$\degree$ toward the $X, Y, Z$ directions. We chose the zigzag antiferromagnetic system since it is known from experiments~\cite{parkEmergenceIsotropicKitaev2024, caoLowtemperatureCrystalMagnetic2016} and our calculation that the magnetic ground state of $\alpha$-RuCl$_3$ is zigzag antiferromagnetic. For example, if the $\jeff$ moment had initially been aligned along the $X$ direction, the moment was canted by 1$\degree$ toward the $Y$ or $Z$ direction. We computed the sum of the total spin and orbital magnetic moments, that is, the total magnetic moment, before and after canting, and took their difference. We assume that the magnitudes of the spin and orbital magnetic moments remain the same during canting. Then, we can obtain the magnitudes of the fixed spin and orbital magnetic moments for the Ru atom from the sum of moments in the zigzag antiferromagnetic unit cell after canting, because the total spin and orbital moments (of an antiferromagnetic material) are zero before canting.

In short, when the $\jeff$ moments are canted by angle $\Delta \theta$, the $g$ factors are computed from the following equations:
% \begin{equation}
%     \vb{S}^{\rm cant}_i = \frac{\Delta \langle \hat{\vb{S}} \rangle^{\rm tot}}{N_{\rm Ru} \Delta \theta}, \,
%     \vb{L}^{\rm cant}_i = \frac{\Delta \langle \hat{\vb{S}} \rangle^{\rm tot}}{N_{\rm Ru} \Delta \theta},
% \end{equation}
\begin{equation}
    g_{\mu\nu}^{[\xi]}
    = \frac{\Delta M_{{\rm tot}, \mu}^{[\xi]}}{\Delta s^{\rm eff}_{{\rm tot},\nu}}
\end{equation}
and
\begin{align}
    g_{X} &= \frac{g_{XX}^{[\text{zAFM-}Y]} + g_{XX}^{[\text{zAFM-}Z]}}{2}, \nnnl
    g_{Y} &= \frac{g_{YY}^{[\text{zAFM-}X]} + g_{YY}^{[\text{zAFM-}Z]}}{2}, \nnnl
    g_{Z} &= \frac{g_{ZZ}^{[\text{zAFM-}X]} + g_{ZZ}^{[\text{zAFM-}Y]}}{2},
\end{align}
% \begin{align}
%     g_{X} &= \frac{1}{2N_{\rm Ru}|\vseff_i|} \left [ \frac{ \Delta_X M_{X, \rm tot}^{[Y]}}{\Delta\theta } + \frac{ \Delta_X M_{X, \rm tot}^{[Z]}}{\Delta\theta} \right ]\nnnl
%     g_{Y} &= \frac{1}{2N_{\rm Ru}|\vseff_i|} \left [ \frac{ \Delta_Y M_{Y, \rm tot}^{[X]}}{\Delta\theta} + \frac{ \Delta_Y M_{Y, \rm tot}^{[Z]}}{\Delta\theta} \right ]\nnnl
%     g_{Z} &= \frac{1}{2N_{\rm Ru}|\vseff_i|} \left [ \frac{ \Delta_Z M_{Z, \rm tot}^{[X]}}{\Delta\theta} + \frac{ \Delta_Z M_{Z, \rm tot}^{[Y]}}{\Delta\theta} \right ],
% \end{align}
where [$\text{zAFM-}\nu$] denotes the zigzag antiferromagnetic configuration with the effective spins aligned along the $\nu$ direction. $\vb{M}^{[\text{zAFM-}\nu]}_{\rm tot} = \vb{L}^{[\text{zAFM-}\nu]}_{\rm tot} + 2\vb{S}^{[\text{zAFM-}\nu]}_{\rm tot}$, $\Delta_\mu \vb{S}^{[\text{zAFM-}\nu]}_{\rm tot}$ and $\Delta_\mu \vb{L}^{[\text{zAFM-}\nu]}_{\rm tot}$ are respectively the variations in the total spin and orbital moments, originally pointing to the $\nu$ direction, upon canting toward the $\mu$ direction ($\mu,\nu=X,Y,Z$),
$\Delta s^{\rm eff}_{{\rm tot},\nu} = \sum_{i=1}^{N_{\rm Ru}} \Delta s^{\rm eff}_{i,\nu} = N_{\rm Ru} \Delta s^{\rm eff}_{1,\nu}$,
and $N_{\rm Ru} = 4$, the number of Ru atoms in the zigzag antiferromagnetic unit cell [Fig.~\ref{fig:configurations}(a)] used in the calculation. For canting by a small angle $\Delta \theta$ toward the $\nu$ direction, $|\Delta s^{\rm eff}_{{\rm tot},\nu}| = N_{\rm Ru} |\vseff_i| \Delta \theta = 2 \Delta \theta$ since $N_{\rm Ru}=4$ and $|\vseff_i|=\frac{1}{2}$.
These quantities are explained in Fig.~\ref{fig:canting}. $\vb{S}^{[\text{zAFM-}\nu]}_{\rm tot}$ was computed from the following formula:
\begin{equation}
    \vb{S}_{{\rm tot}}^{[\text{zAFM-}\nu]} = 
    \frac{1}{2}
    \sum_{v\vb{k}}^{\rm occ}
    \int
    \langle \psi_{v\vb{k}}^{[\text{zAFM-}\nu]}|\vb{r}\sigma\rangle
    \boldsymbol{\sigma}_{\sigma\sigma'}
    \langle \vb{r}\sigma'|\psi_{v\vb{k}}^{[\text{zAFM-}\nu]} \rangle \,d\vb{r} \,.
    % \mel{\psi_{v\vb{k},\nu}}{\hat{\vb{S}}}{\psi_{v\vb{k},\nu}} \, ,
\end{equation}
For the computation of $\vb{L}^{[\text{zAFM-}\nu]}_{\rm tot}$, we used the modern theory of orbital magnetization in Refs.~\onlinecite{2006orb_magnetization, 2012orbital_magnetization}.
% After this determination of the spin and orbital moments, we compute the magnitude of the total magnetic moment from the sum of the spin and orbital moments and multiply by 2 to obtain the $g$ factor for the $J_\text{eff}=1/2$ moment.
% Finally, we symmetrized the $g$ factor by the $C_3$ rotation.
% since we want the model Hamiltonian to be $C_3$ symmetric.
To compute total orbital magnetization, we used the Wannier interpolation~\cite{2012orbital_magnetization} and the translation-equivariant formulas~\cite{transl_inv}.
We note that the translation-equivariant formulas are necessary for effective and accurate computation of orbital magnetization.

Also, we compared the Wannier interpolated results with the results obtained from projection to the Hubbard manifold. For the atomic orbital projection method, we used the Hubbard occupation matrix $\rho_{i,mn\sigma\sigma'}$ for atom $i$.
% The spin and orbital angular momentum operator $\hat{\vb{S}}$ and $\hat{\vb{L}}$ are represented by the $10\times10$ matrices $\vb{S}_{i,nm} = \mel{\phi_{i,n}}{\hat{\vb{S}}}{\phi_{i,m}}$ and $\vb{L}_{i,nm} = \mel{\phi_{i,n}}{\hat{\vb{L}}}{\phi_{i,m}}$, respectively, in the spin-$d$-orbital basis of atom $i$.
Then, we can compute the $g$ factors by using the atomic-orbital projected moments:
\begin{equation}
    M_{i,\mu}^{[\text{zAFM-}\nu]} = g_{\mu\nu} s^{{\rm eff}}_{i,\nu}
\end{equation}
and
\begin{align}
    g_{X} = \frac{M_{i,X}^{[\text{zAFM-}X]}}{s^{{\rm eff}}_{i,X}},\,
    g_{Y} = \frac{M_{i,Y}^{[\text{zAFM-}Y]}}{s^{{\rm eff}}_{i,Y}},\,
    g_{Z} = \frac{M_{i,Z}^{[\text{zAFM-}Z]}}{s^{{\rm eff}}_{i,Z}},
\end{align}
where $\vb{M}^{[\text{zAFM-}\nu]}_{i} = \vb{L}^{[\text{zAFM-}\nu]}_{i} + 2 \vb{S}^{[\text{zAFM-}\nu]}_{i}$.
%and $\vb{L}^{[\mu]}_{i}$ and $\vb{S}^{[\mu]}_{i}$ are respectively the orbital and spin moments obtained from atomic projection for the zigzag anti-ferromagnetic system with the collinear direction $\mu$. 
% \begin{align}
%     \vb{L}^{[\mu]}_{i} &= \sum_{mn\sigma\sigma'} \rho_{i,mn\sigma\sigma'}^{[\mu]} \vb{L}_{nm\sigma'\sigma} \nnnl
%     \vb{S}^{[\mu]}_{i} &= \sum_{mn\sigma\sigma'} \rho_{i,mn\sigma\sigma'}^{[\mu]} \vb{S}_{nm\sigma'\sigma} ,
% \end{align}
% where
% \begin{equation}
%     \rho_{i,mn\sigma\sigma'}^{[\mu]} = \sum_{v\vb{k}}^{\rm occ} \langle \phi_{i,m\sigma} | \psi_{v\vb{k}}^{[\mu]} \rangle \langle \psi_{v\vb{k}}^{[\mu]} | \phi_{i,n\sigma'} \rangle .
% \end{equation}
After the determination of the $g$ factors, we symmetrized the in-plane components by using the $C_3$ rotation, defining $g_{XY} = (g_X + g_Y)/2$. Note that the off-diagonal part of the anisotropic $g$ factor is found to be negligible according to our calculations, and thus, it suffices to take average of $X$ and $Y$ components of $g$ factor for $C_3$ symmetrization.
% $\vb{S}^{\rm atom}_i = \sum_{mn}  \rho_{i,mn} \vb{S}_{i,nm}$ and $\vb{L}^{\rm atom}_i = \sum_{mn}  \rho_{i,mn} \vb{L}_{i,nm}$.
% Then, we can compute the atomic-orbital projected moments by $\vb{S}^{\rm atom}_i = \Tr \left ( \hat{\rho}^i \hat{\vb{S}}^i \right )$ and $\vb{L}^{\rm atom}_i = \Tr \left ( \hat{\rho}^i \hat{\vb{L}}^i \right )$.

Finally, to study the effects of structural distortion from a perfect RuCl$_6$ octahedron on the magnetism of the monolayer, we computed the total energy of the zigzag antiferromagnetic state as a function of the effective spin axis direction. The collinear, effective spin direction was confined and varied within the $XZ$ plane, where the lowest energy configuration can be found for a given structural distortion. In the case of trigonal distortion, we varied the $Z$ coordinate of the Cl atoms and fixed their $XY$ coordinates to the experimental structure so that the angle between the $Z$ and $z$ axes is equal to that of a perfect octahedron, $\cos^{-1} (1/\sqrt{3}) = 54.74 \degree$. In the case of the relative twist of the upper and lower Cl triangles, we varied the $XY$ coordinates of the Cl atoms and fixed their $Z$ coordinates to the experimental ones so that the $XY$ coordinates are the same as those of a perfect octahedron. We examined and compared the effects of the two types of distortions.

\section{Computational Details}
We used our custom version of \texttt{Quantum ESPRESSO} for the constrained DFT+$U$ calculation of $\alpha$-RuCl$_3$~\cite{giannozzi_quantum_2009,giannozzi_advanced_2017}. We employed the local density approximation (LDA) from Ref.~\onlinecite{PZ_functional} for the exchange-correlation functional and the rotationally invariant LDA+$U$ method of Ref.~\onlinecite{PhysRevB.52.R5467}. We used $U=2,\,3,\,4$~eV for the Hubbard parameters of Ru. Norm-conserving full-relativistic pseudopotentials generated from the optimized norm-conserving Vanderbilt pseudopotential (\texttt{ONCVPSP}) suite were used for both Ru and Cl~\cite{scherpelz_implementation_2016, PhysRevB.88.085117, SCHLIPF201536}. For pseudopotential generation, we used the parameters from \texttt{PseudoDojo}~\cite{2018vanSettenPseudodojo}. %We used the LDA since the non-collinear generalized gradient approximation has serious convergence problems as discussed in Ref.~\onlinecite{kimMagneticAnisotropyMagnetic2021}.
We did not use a generalized gradient approximation (GGA) functional in our computation since, as described in Ref.~\onlinecite{kimMagneticAnisotropyMagnetic2021}, the use of the noncollinear version of GGA for magnetic systems results in serious convergence problems. Because this convergence issue significantly affect the accuracy of the DFT total energies, which needed to be determined within a few hundred \textmu{}eV, we decided not to use the noncollinear GGA in our calculations. Instead, we also performed a fixed-cell relaxation, where the lattice constants were fixed to the experimental values. Our results show that the $R\bar{3}$ structure is lower in energy also in the fixed-cell relaxation, demonstrating that our result does not change with lattice constants. The value of the constraining parameter $\Lambda$ [see Eq.~\eqref{eq:Econstr}] was in the range from 0.1 to 1 Ry, depending on the $U$ value. Note that when fully converged, the constraining energy [Eq.~(\ref{eq:Econstr})] was nearly zero, and therefore the conclusions do not depend on the specific value of $\Lambda$. For relaxation, the kinetic energy cutoff was set to 75 Ry and the $k$-point grid was set to $8 \times 8\times 8$ for both $C2/m$ and $R\bar{3}$ bulk structures. %Also, we performed both variable-cell relaxation and fixed-cell relaxation with experimental lattice constants to check the influence of the lattice constants.% This is because LDA has a tendency to underestimate the lattice constants.

We found that a proper initialization of the Hubbard occupation matrix is essential to achieve convergence. The initial occupation matrix was constructed with the $\jeff = 3/2$ and $\jeff = 1/2$ orbitals with respect to the local octahedron $xyz$ axes. We set the occupation of $\jeff = 3/2$ orbitals to one. Then, we constructed a linear combination of the two $J_\text{eff} = 1/2$ states to obtain a state with all $\vseff_i$ moments pointing to desired directions. This state was set to be occupied by a single electron in the initial Hubbard occupation matrix. The self-consistent field convergence threshold was set to $10^{-10}$ Ry. For structural relaxation, the energy convergence threshold was $10^{-5}$ Ry, while the force convergence threshold was set to $ 2.5\times 10^{-4}$ Ry/Bohr. The stress convergence threshold was set to 0.25 kbar. 

In extracting the anisotropic exchange parameters, we used the experimental $R\bar{3}$ structure. We did not use the $C2/m$ structure since the ground-state stacking is $R\bar{3}$-type according to recent experiments~\cite{parkEmergenceIsotropicKitaev2024, RuCl3_R3, RuCl3_R3_2, RuCl3_R3_3} and our calculations. The monolayer structure was obtained from the bulk structure by taking a single layer therefrom. We note that since the monolayer slab of the experimental bulk structure was used in our calculations, the resulting monolayer properties are expected to more closely reflect those of the bulk, provided that the interlayer exchange interactions are much smaller than the intralayer exchange interactions. \citet{kim_crystal_2016} estimated the magnitude of the interlayer exchange coupling in this material and concluded that it is negligible ($\sim$ 0.05 meV) compared to the energy scale of the other intralayer couplings. The kinetic cutoff was 50 Ry, and the $k$-point grid was $6 \times 6  \times 1$ for the monolayer primitive cell with two Ru atoms.

Finally, for the computation of orbital magnetization, we used the {\texttt{Wannier90}}~\cite{pizziWannier90CommunityCode2020,PhysRevB.56.12847,PhysRevB.65.035109} and our translation-equivariant~\cite{transl_inv} version of the {\texttt{Wannierberri}}~\cite{tsirkinHighPerformanceWannier2021a} packages. The coarse \textit{ab initio} grid for the Wannier interpolation was $12 \times 6 \times 4$ and the fine grid was $120 \times 60 \times 40$. Note that to accurately compute the orbital magnetic moments, we needed fine sampling along the vacuum direction even in the two-dimensional slab geometry calculation~\cite{transl_inv}. The number of bands used for the Wannier interpolation was 168, and the number of Wannier functions is the same, since the interpolated bands were isolated from the others. We used the Ru $s,p,d$ and Cl $s,p$ orbitals for the initial projection. Projection-only Wannier functions were used for interpolation.

\section{Results and Discussion}

%We first examine the results of relaxation. The results for $U=3$ eV are shown in Table~\ref{table:relaxation}.
\subsection{Bulk stacking: $C2/m$ vs $R\bar{3}$}

\begin{table}[ht]
    \centering
    \caption{Total energy of bulk $\alpha$-RuCl$_3$ in $C2/m$ and $R\bar{3}$ structures per Ru atom obtained with experimental structure~\cite{parkEmergenceIsotropicKitaev2024}, fixed-cell (fixing the lattice parameters to the experimentally measured ones) and variable-cell structural relaxations. The energy values are in units of meV.}
    \begin{tblr}{
        cells = {c},
        hline{1,5,9,13} = {-}{0.08em},
        hline{2,6,10} = {-}{0.04em},
        vline{2} = {-}{0.04em}
    }
    $U=2$ eV & Experimental & Fixed-cell & Variable-cell \\
    $C2/m$ & 46.53 & 23.68 & 5.64 \\
    $R\bar{3}$ & 28.14 & 17.28 & 0.00 \\
    $\Delta E$ & 18.39 & 6.40 & 5.64 \\
    $U=3$ eV & Experimental & Fixed-cell & Variable-cell \\
    $C2/m$ & 39.17 & 19.96 & 5.29 \\
    $R\bar{3}$ & 22.39 & 13.81 & 0.00 \\
    $\Delta E$ & 16.78 & 6.15 & 5.29 \\
    $U=4$ eV & Experimental & Fixed-cell & Variable-cell \\
    $C2/m$ & 33.63 & 17.84 & 5.04 \\
    $R\bar{3}$ & 18.46 & 11.94 & 0.00 \\
    $\Delta E$ & 15.17& 5.90 & 5.04
    \end{tblr} \label{table:relaxation}
\end{table}

Table~\ref{table:relaxation} shows that the $R\bar{3}$ structure is lower in energy than the $C2/m$ structure in all cases: the experimental structure, variable-cell relaxation, and fixed-cell relaxation. The energy difference is about 6.15 meV/Ru for the fixed-cell relaxation, in which the lattice parameters were set to the measured ones and only the atomic positions were relaxed, and 5.29 meV/Ru for the variable-cell relaxation, in which both the lattice parameters and atomic positions were relaxed. Therefore, our calculation results confirm for the first time the recent experimental reports that the $R\bar{3}$ structure is more stable than the $C2/m$ structure~\cite{parkEmergenceIsotropicKitaev2024, RuCl3_R3, RuCl3_R3_2, RuCl3_R3_3}. As shown in Table~\ref{table:relaxation}, the same conclusion holds for different Hubbard $U$ parameters.

\begin{table*}[ht]
  \centering
  \caption{Experimentally measured structure of bulk $\alpha$-RuCl$_3$~\cite{parkEmergenceIsotropicKitaev2024} and relaxed $R\bar{3}$ structures obtained with $U=3$ eV. The trigonal angle is defined as the angle between Ru-Cl bond and the $Z$ axis.}
  \begin{tblr}{
      cells = {c},
      hline{1,10} = {-}{0.08em},
      hline{2} = {-}{0.04em},
      vline{2} = {-}{0.04em}
  }
  & Experiment & Fixed-cell relaxation & Variable-cell relaxation\\
  $a$ (\AA) & 5.973 & 5.973 &  5.912 \\
  $b$ (\AA) & 11.946 & 11.946 & 11.825 \\
  $c$ (\AA) & 16.930 & 16.930 & 16.378 \\
  $\alpha$ ($\degree$) & 90.00 & 90.00 & 90.02 \\
  $\beta$ ($\degree$) & 90.00 & 90.00 & 89.93 \\
  $\gamma$ ($\degree$) & 120.00 & 120.00 & 120.01 \\
  Average Ru-Cl bond length (\AA) & 2.356 & 2.341 & 2.336 \\
  Trigonal angle ($\degree$)& 55.53 & 56.03 & 55.87
  \end{tblr} \label{table:relaxation_details}
\end{table*}

We now examine the details of the relaxed structures presented in Table~\ref{table:relaxation_details}. After the variable-cell relaxation, the lattice constants show a slight contraction (by 1.02\%) from those of the experimental structure. This is expected since the LDA functional tends to underestimate lattice constants. We also note that because of the zigzag antiferromagnetic order, the lattice angles were also slightly altered (by 0.02$\degree$, 0.07$\degree$, and 0.01$\degree$ for $\alpha,\beta$ , and $\gamma$, respectively) from their corresponding experimental values. The average Ru-Cl bond length was also slightly reduced (by 0.64\% and 0.85\% for fixed-cell and variable-cell relaxation, respectively) from the experimental structure in both relaxations. Finally, the trigonal angle, which is defined as the angle between the Ru-Cl bond and the $Z$ axis, increased slightly (by 0.50$\degree$ and 0.34$\degree$ for fixed-cell and variable-cell relaxation, respectively) from the measured value in both relaxations. Since the trigonal angle for an ideal octahedron is $54.74 \degree$, relaxation results in a slightly more trigonally distorted octahedron.

\subsection{Electronic structure: $J_\textrm{eff}$ picture}

\begin{figure}[ht!]
\label{fig:PDOSJeff}
\centering
\includegraphics[width=1.0\linewidth]
{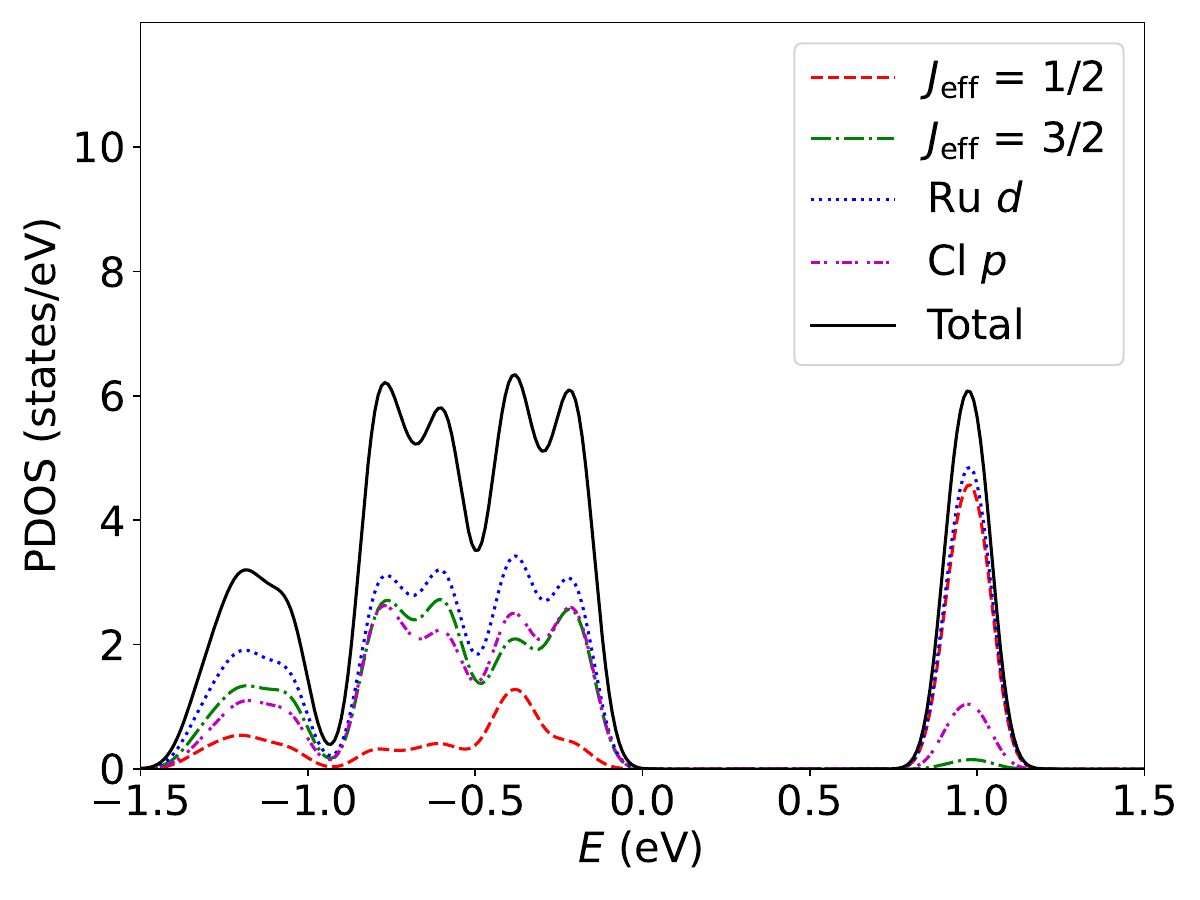}
\caption{
Projected density of states (PDOS) of $\alpha$-RuCl$_3$ around the valence band maximum ($E = 0$~eV) and the conduction band minimum ($E = 0.8$~eV).
}
\label{fig:pdos}

\end{figure}

\begin{figure}[ht!]
\centering
\includegraphics[width=1.0\linewidth]
{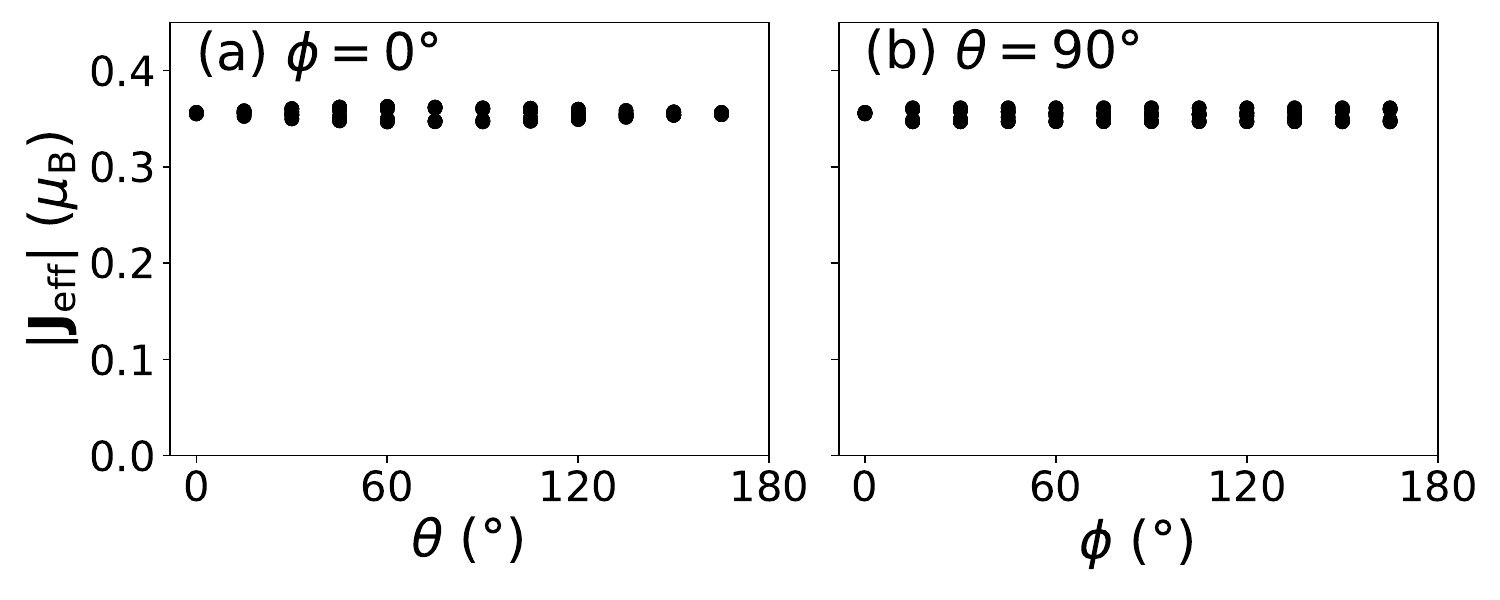}
\caption{
Magnitude of $\mathbf{J}_\text{eff}$ moments of all four or eight Ru atoms in a unit cell in all 60 magnetic configurations. In total, there are $4\times 48 + 8 \times 12=288$ symbols (48 configurations with four Ru ions per unit cell and 12 configurations with 8 Ru ions per unit cell) in each panel. $\theta$ and $\phi$ represent the collinear moment direction $\vb{e}^{[\xi]} = (\sin\theta \cos\phi, \sin\theta \sin\phi, \cos\theta)^{\mathsf{T}}$.
}
\label{fig:jeff_magnitude}
\end{figure}

To demonstarte that $\alpha$-RuCl$_3$ is described by the $J_\text{eff}$ picture, we present the total density of states (DOS) and the projected density of states (PDOS) in Fig.~\ref{fig:pdos}. In the valence band maximum ($E = 0$~eV), the total DOS is composed of similar contributions from the Ru $d$ orbitals and Cl $p$ orbitals. The Ru $d$ orbitals are well explained by the $J_\text{eff} = 3/2$ and $J_\text{eff} = 1/2$ states. These results show that the electronic states of $\alpha$-RuCl$_3$ near the band gap are explained by mixed $J_\text{eff} = 3/2$ and $J_\text{eff} = 1/2$ states with the hybridization of Cl $p$ states. Also, the conduction band minimum is well explained by $J_\text{eff} = 1/2$ states with a small contribution of Cl $p$ states. These results are in agreement with \citet{RuCl3_jeff_first}.

We also verified the assumption in Eq.~\eqref{eq:def_s_eff} by computing $\abs{\mathbf{J}_\text{eff}}$ of all four or eight Ru atoms in a unit cell in all configurations, i.\,e.\, in total $4\times 48 + 8 \times 12 = 288$ cases (48 configurations with 4 Ru ions per unit cell and 12 configurations with 8 Ru ions per unit cell). Figure~\ref{fig:jeff_magnitude} shows that the magnitude of $\mathbf{J}_\text{eff}$ is indeed a constant, to a good approximation, across all configurations.
In passing, we note that $|\mathbf{J}_i^\text{eff}|$ is not equal to 1/2, which is the value obtained from the ideal $J_\text{eff}$ picture, but of a smaller magnitude, because the occupations of the two $J_\text{eff}$ states are not equal to 1 or 0. This discrepancy from the ideal $J_{\text{eff}}$ model is also reported by \citet{houUnveilingMagneticInteractions2017a}. There are several reasons for this difference. The atomic projection cannot capture the hybridization of Ru $d$ states with Cl $p$ states, causing the system to deviate from the ideal $J_\text{eff}$ picture. Also, distortions from a perfect octahedral structure mix $J_\text{eff} = 3/2$ and $J_\text{eff} = 1/2$ states. Lastly, the spin-orbit coupling might not be strong enough to lead to the perfect $J_\text{eff}$ picture.

Thus, caution should be taken in the normalization of moments when mapping the model magnetic Hamiltonian to the DFT results. Therefore, further research would be needed on the normalization of the effective moments during the mapping to the DFT total energies.

%\subsection{$J_\textrm{eff}$ vs. $L_\textrm{eff}$- and $S$-polarized basis states}

\begin{figure*}[ht!]
    \centering
    \includegraphics[width=0.95\linewidth]{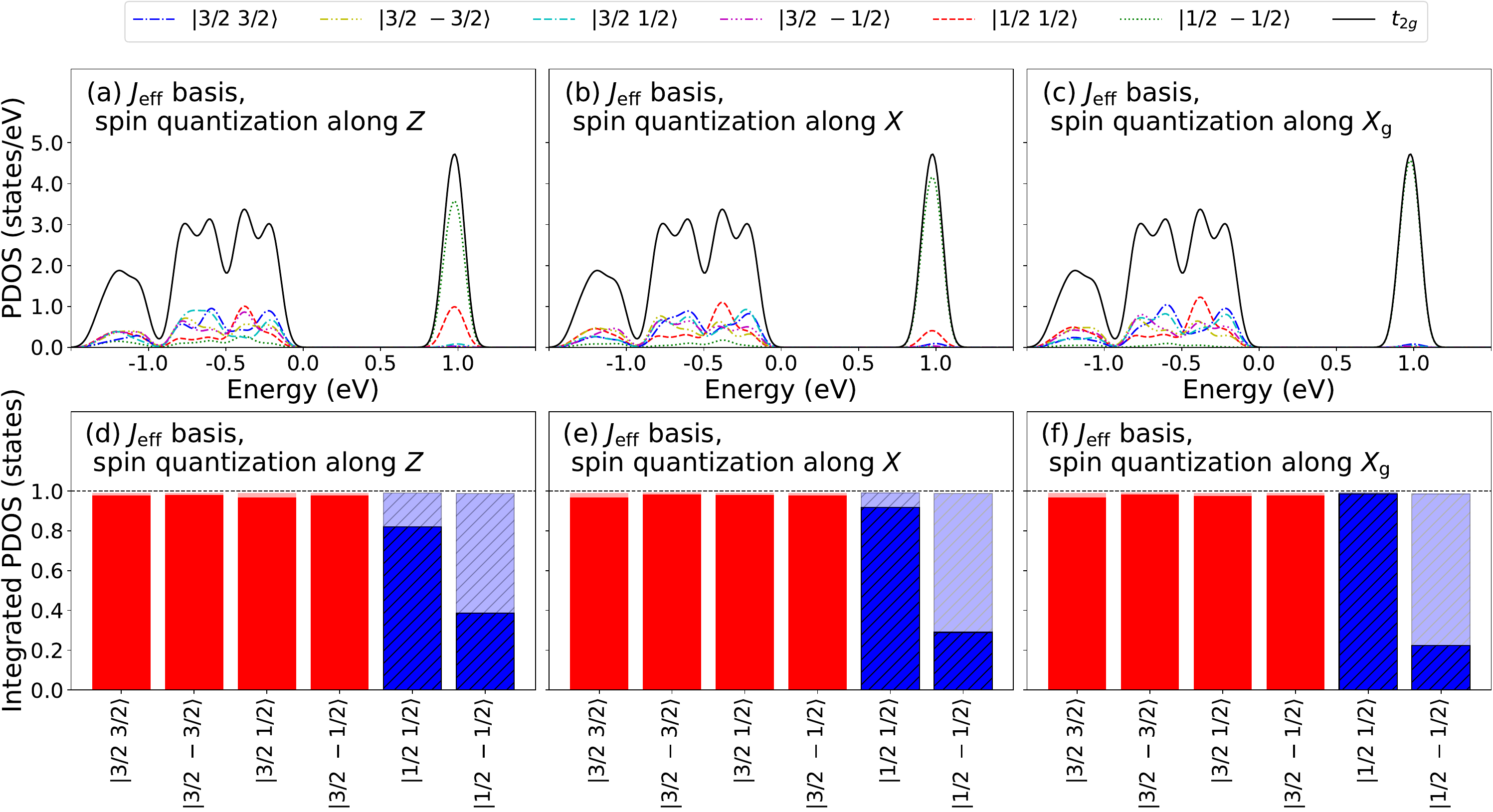}
    \caption{
PDOS and integrated PDOS computation results. (a)-(c) show the PDOS's using the $J_\text{eff}$ basis states, while (d)-(f) show the integrated PDOS's. For (d), (e), and (f), the lower filled bars and the upper faint bars correspond to the integrated PDOS below and above the band gap, respectively. The $X_\text{g}$ direction corresponds to the N\'eel vector direction, that is, $(\theta, \phi) = (55 \degree , 0 \degree)$.}
    \label{fig:PDOS_analysis}
\end{figure*}

To further examine the $J_\text{eff}$ picture, we computed the PDOS using the $m-$resolved $J_\text{eff}$ basis states. We also computed the integrated PDOS below and above the band gap for each of the basis states.

% {\color{red}
% \sout{
% Specifically, we used two sets of Hubbard parameters for DFT$+U$ computation: $(U, J_\text{H})=$(3, 0)~eV and $(U, J_\text{H})=$(2, 0.5)~eV, where $J_\textrm{H}$ is the intra-atomic Hund's coupling. For each parameter set, we performed two types of computation. For the first one, we initialized the Hubbard occupation matrix following the $J_\text{eff}=1/2$ picture and constrained the $J_\text{eff}$ moments. For the second one, we initialized the occupation matrix following the $L_\textrm{eff}$- and $S$-polarized picture and constrained the spin moments because this basis is composed of eigenstates of both spin and (effective) orbital angular momentum operators. The constraint method for the spin moments is almost the same as the constraint method for the $J_{\rm eff}$ moments.
% }
% }
For all computations, the constraining configuration was set to the magnetic ground state configuration whose N\'eel vector is within the $XZ$ plane and is close to the $X$ axis [$(\theta, \phi) = (55 \degree , 0 \degree)$ in spherical coordinates], which is in agreement with the experiment~\cite{searsFerromagneticKitaevInteraction2020, zigzag_conf}. Moreover, we show the results by projecting the wave functions on the basis states whose quantization axis is either along the $Z$ direction, the $X$ direction, or along the N\'eel vector direction obtained from our calculations ($X_\textbf{g}$) [$(\theta, \phi) = (55 \degree , 0 \degree)]$.

Figure~\ref{fig:PDOS_analysis} shows the PDOS and integrated PDOS results.Figures~\ref{fig:PDOS_analysis} (a),~\ref{fig:PDOS_analysis}(b), and~\ref{fig:PDOS_analysis}(c) show that the conduction band minimum  is almost $J_\text{eff}=1/2$ states. Especially, Figure~\ref{fig:PDOS_analysis}(c) demonstrates that the conduction band minimum is almost explained by the $J_\text{eff}=1/2$ state with the $J_\text{eff}$ moment direction being parallel to the N\'eel vector.
The difference among Figs.~\ref{fig:PDOS_analysis}(a),~\ref{fig:PDOS_analysis}(b), and~\ref{fig:PDOS_analysis}(c) shows that the appropriate choice of the spin quantization axis, which has so far been neglected in the study of RuCl$_3$, is very important to fully understand the orbital character of the valence and conduction band states.

To quantify the occupation of each basis state, we calculated the integrated PDOS below and above the band gap of each basis state
[Figs.~\ref{fig:PDOS_analysis}(d),~\ref{fig:PDOS_analysis}(e), and~\ref{fig:PDOS_analysis}(f)].
Especially, Fig.~\ref{fig:PDOS_analysis}(f) shows that only the $\ket{1/2 \,\, -1/2}$ state is (almost) empty while the others are fully occupied if the quantization axis is chosen to be along the N\'eel vector direction, strongly supporting the $J_\text{eff}$ scenario.
%For the $J_\text{eff}$ basis with the spin quantization axis along the N\'eel vector direction, only the $\ket{1/2 \,\, -1/2}$ state (second blue bar in Fig.~\ref{fig:PDOS_initial_Jeff_constrain_Jeff_U_3} (d)) shows large integrated PDOS above the band gap, which supports that the conduction band minimum is almost $J_\text{eff}=1/2$ character.
Also, the difference among Figs.~\ref{fig:PDOS_analysis}(d),~\ref{fig:PDOS_analysis}(e), and~\ref{fig:PDOS_analysis}(f) shows, again, that the appropriate choice of the spin quantization axis is essential to fully understand the orbital character of the conduction band minimum.

We performed similar analysis for a different set of Hubbard parameters $(U, J_\text{H})=$(2, 0.5)~eV following \citet{Liu2023}, where $J_\textrm{H}$ is the intra-atomic Hund's coupling, tried the $L_\textrm{eff}$ and $S$-polarized initial Hubbard occupation matrix, and constrained the direction of $S$ polarization. From these several calculations, we learned that the $J_\textrm{eff}$ state we obtained is quite robust and does not depend on the Hubbard parameters, initial Hubbard occupation matrix, or the constraining scheme.% within the fully self-consistent treatment of relativistic effects, i.\,e.\,, is robust with respect to various constraining DFT and spin-orbit coupling implementations.% These results are in agreement with \citet{RuCl3_jeff_first}.

\subsection{Magnetic exchange parameters}

\begin{figure*}[t]
    \centering
    \includegraphics[width=1.0\textwidth]{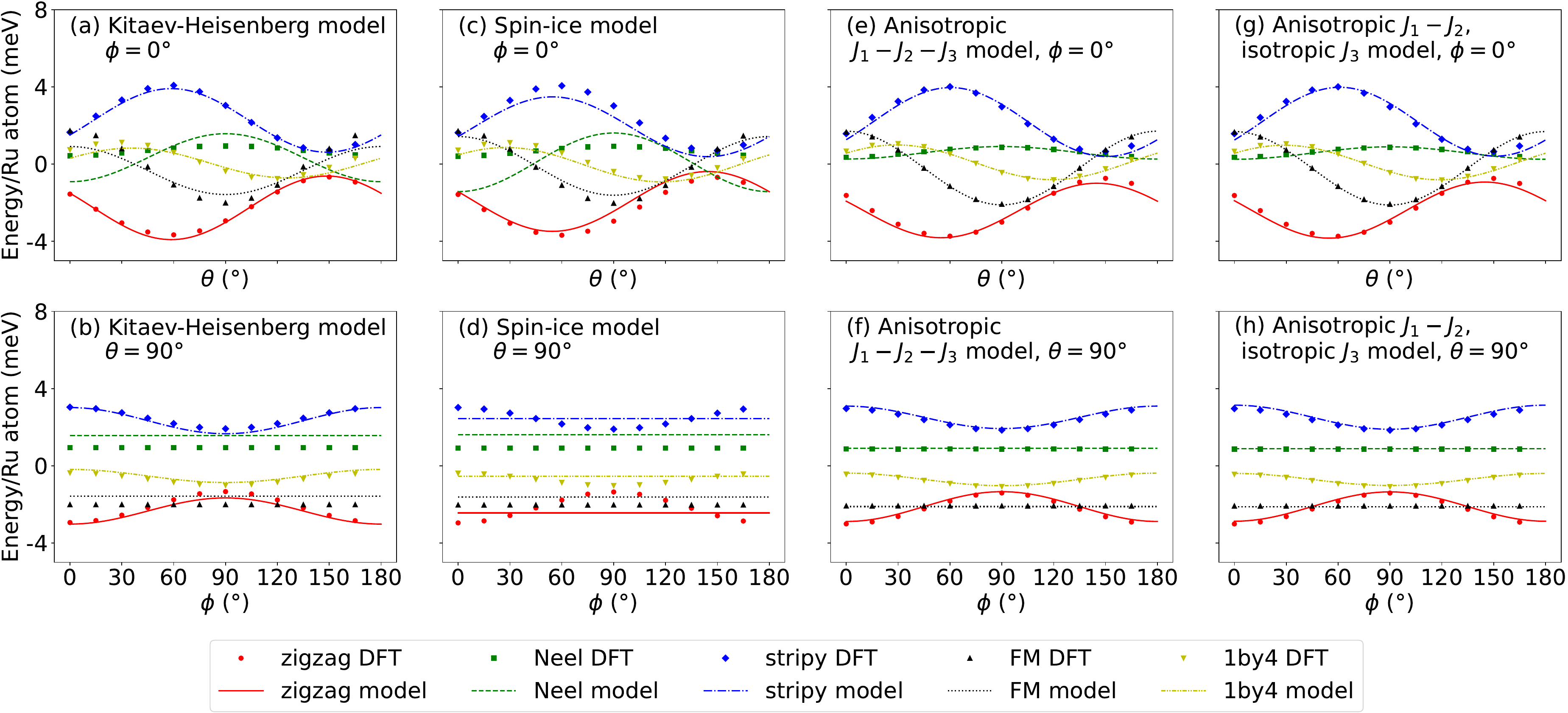}
    \caption{The total energy versus the effective moment direction of monolayer $\alpha$-RuCl$_3$ obtained from DFT calculations (symbols) and from the models made from the DFT calculations (curves). The DFT calculations were done on the experimentally obtained $R\bar{3}$ structure and with $U=3$ eV. $\theta$ and $\phi$ represent the collinear moment direction $\vb{e}^{[\xi]} = (\sin\theta \cos\phi, \sin\theta \sin\phi, \cos\theta)^{\mathsf{T}}$.}
    \label{fig:fitting}
\end{figure*}

We now discuss the magnetic exchange parameters of the monolayer, summarized in Table~\ref{table:parameters}.
%We note that though the overall magnitude of exchanges is different for different $U$, the qualitative features are the same. 
Figure~\ref{fig:fitting} compares the computed DFT total energies and fitted model energies for each configuration in Fig.~\ref{fig:configurations} with the collinear direction $\vb{e}^{[\xi]} = (\sin\theta \cos\phi, \sin\theta \sin\phi, \cos\theta)^{\mathsf{T}}$.
As shown in Figs.~\ref{fig:fitting}(a) and (b), the conventional KH model inaccurately describes the DFT total energies, particularly for the N\'eel configuration.

In the case of the SI model [Figs.~\ref{fig:fitting}(c) and (d)], the discrepancy between the model and the DFT results becomes even larger.  For the $\theta=90\degree$ plane [Fig.~\ref{fig:fitting}(d)], the DFT energy is significantly dependent on $\phi$, the azimuthal angle of the $\jeff${} moments, while the model energy is completely independent of $\phi$. This is an inherent feature of the SI model. For example, in the zigzag configuration where the same-spin zigzag chain runs along the $X$ direction, the nearest-neighbor contribution, which is equal to the total contribution, to the total magnetic anisotropy tensor is written as
\begin{equation}
\aj_{\text{zigzag AFM}}^{\text{tot}} = R \aj^{\rm SI}_{1,ij} R^{-1} + R^{-1} \aj^{\rm SI}_{1,ij} R - \aj^{\rm SI}_{1,ij},
\end{equation}
where $R$ is the $120 \degree$ rotation matrix [Eq.~(\ref{eq:rot_120})] and $\aj^{\rm SI}_{1,ij}$ is the nearest-neighbor exchange interaction tensor [Eq.~\eqref{eq:model2}] between the two Ru atoms connected along the $Y$ direction.
A direct calculation reveals that the above tensor leads to the total energy being isotropic for the rotation of the effective spin direction in the $XY$ plane:
\begin{widetext}
\begin{align}
\begin{split}
\aj_{\text{zigzag AFM}}^{\text{tot}} &=
\mqty(-1/2 & -\sqrt{3}/2 & 0 \\ \sqrt{3}/2 & -1/2 & 0 \\ 0 & 0 & 1)
\mqty(J_{1,XX} & 0 & J_{1,ZX} \\ 0 & J_{1,XX} & 0 \\ J_{1,ZX} & 0 & 0)
\mqty(-1/2 & \sqrt{3}/2 & 0 \\ -\sqrt{3}/2 & -1/2 & 0 \\ 0 & 0 & 1) 
\\ &+
\mqty(-1/2 & \sqrt{3}/2 & 0 \\ -\sqrt{3}/2 & -1/2 & 0 \\ 0 & 0 & 1)
\mqty(J_{1,XX} & 0 & J_{1,ZX} \\ 0 & J_{1,XX} & 0 \\ J_{1,ZX} & 0 & 0)
\mqty(-1/2 & -\sqrt{3}/2 & 0 \\ \sqrt{3}/2 & -1/2 & 0 \\ 0 & 0 & 1) -
\mqty(J_{1,XX} & 0 & J_{1,ZX} \\ 0 & J_{1,XX} & 0 \\ J_{1,ZX} & 0 & 0)
\\ &=
\mqty(J_{1,XX} & 0 & -J_{1,ZX}/2 \\ 0 & J_{1,XX} & \sqrt{3}J_{1,ZX}/2 \\ -J_{1,ZX}/2 & \sqrt{3}J_{1,ZX}/2 & 0)
+
\mqty(J_{1,XX} & 0 & -J_{1,ZX}/2 \\ 0 & J_{1,XX} & -\sqrt{3}J_{1,ZX}/2 \\ -J_{1,ZX}/2 & -\sqrt{3}J_{1,ZX}/2 & 0) -
\mqty(J_{1,XX} & 0 & J_{1,ZX} \\ 0 & J_{1,XX} & 0 \\ J_{1,ZX} & 0 & 0)
\\ &=
\mqty(J_{1,XX} & 0 & -2J_{1,ZX} \\ 0 & J_{1,XX} & 0 \\ -2J_{1,ZX} & 0 & 0)\,,
\end{split}
\end{align}
and hence the contribution to the total energy with the collinear moment direction $\vb{e}^{[\xi]} = (\sin\theta \cos\phi, \sin\theta \sin\phi, \cos\theta)^{\mathsf{T}}$ is
\begin{align}
\begin{split}
    \vb{e}^{[\xi]} \cdot \aj_{\text{zigzag AFM}}^{\text{tot}}\, \vb{e}^{[\xi]} &=
    \frac{1}{4} \mqty(\sin\theta \cos\phi & \sin\theta\sin\phi & \cos \theta) \mqty(J_{1,XX} & 0 & -2J_{1,ZX} \\ 0 & J_{1,XX} & 0 \\ -2J_{1,ZX} & 0 & 0)
    \mqty(\sin \theta\cos\phi \\ \sin\theta\sin\phi \\ \cos\theta) \\ &=
    \frac{1}{4} \sin \theta \qty(J_{1,XX} \sin \theta - 4J_{1,ZX} \cos \theta \cos \phi)\,,
\end{split}
\end{align}
\end{widetext}
which proves that the model total energy is independent of $\phi$ at $\theta = 90 \degree$, i.e.,
$\vb{e}^{[\xi]} \cdot \aj_{\text{zigzag AFM}}^{\text{tot}}\, \vb{e}^{[\xi]} = (J_{1,XX} \sin^2\theta)/4$.
% This qualitative difference shows that the magnetic properties of $\alpha$-RuCl$_3$ can not be described in terms of the spin-ice model.

In contrast, our proposed A-$J_1J_2J_3$ model accurately reproduces the DFT total energies of all magnetic configurations and collinear moment directions [Figs.~\ref{fig:fitting}(e) and (f)]. % The larger amplitude for the Neel configuration is resolved and the discrepancy of energies for the Neel and ferromagnetic configuration in the $\theta=90\degree$ has disappeared. 

\begin{table*}[ht]
\centering
\caption{Magnetic exchange parameters of monolayer $\alpha$-RuCl$_3$ for the models with $U=3$ eV. All quantities are in units of meV.}
\begin{tblr}{
  cells = {c},
  vline{2} = {-}{},
  hline{1,6,11,16} = {-}{0.08em},
  hline{2,7,12} = {-}{},
}
        & $J_1$ & $K_1$ & $\Gamma_1$ & $\Gamma'_1$ \\
Conventional model & $-3.76 \pm 0.29$ & $-5.57 \pm 0.70$ & $6.78 \pm 0.34$ & $-0.07 \pm 0.21$ \\
Spin-ice model & $-2.64 \pm 0.32$ & $-8.28 \pm 0.63$ & $4.14 \pm 0.27$ & $1.32 \pm 0.16$ \\
Anisotropic $J_1$-$J_2$-$J_3$ model & $-3.88 \pm 0.09$ & $-5.39 \pm 0.24$ & $6.47 \pm 0.12$ & $-0.29 \pm 0.07$ \\ 
Anisotropic $J_1$-$J_2$ isotropic $J_3$ model & $-3.87 \pm 0.08$ & $-5.38\pm 0.20$ & $6.54 \pm 0.10$ & $-0.30 \pm 0.06$ \\ 
& $J_2$ & $K_2$ & $\Gamma_2$ & $\Gamma'_2$ \\
Conventional model & - & - & - & - \\
Spin-ice model & - & - & - & - \\
Anisotropic $J_1$-$J_2$-$J_3$ model & $0.06 \pm 0.05$ & $-0.48 \pm 0.12$ & $0.68 \pm 0.06$ & $0.72 \pm 0.03$ \\ 
Anisotropic $J_1$-$J_2$ isotropic $J_3$ model & - & $-0.34\pm 0.07$ & $0.68 \pm 0.05$ & $0.71 \pm 0.03$ \\ 
        & $J_3$ & $K_3$ & $\Gamma_3$ & $\Gamma'_3$ \\
Conventional model & $3.63 \pm 0.15$ & - & - & - \\
Spin-ice model & $3.80 \pm 0.17$ & - & - & - \\
Anisotropic $J_1$-$J_2$-$J_3$ model & $3.68 \pm 0.09$ & $-0.14 \pm 0.24$ & $-0.07 \pm 0.11$ & $0.06 \pm 0.06$        \\
Anisotropic $J_1$-$J_2$ isotropic $J_3$ model & $3.62 \pm 0.04$ & - & - & -
\end{tblr}
\label{table:parameters}
\end{table*}

The exchange parameters of the three models are shown in Table~\ref{table:parameters}. The anisotropic part of the second-nearest-neighbor exchange parameters, $K_2$, $\Gamma_2$, and $\Gamma'_2$, which have been neglected so far, is non-negligible. Furthermore, the number of second-nearest-neighbor Ru atoms (6) is twice as large as that of the first- or third-nearest-neighbor Ru atoms (3). Therefore, in terms of energetics, the second-nearest-neighbor exchange parameters are even more important.

\citet{yadavKitaevExchangeFieldinduced2016a} examined the effect of isotropic second-nearest-neighbor exchange. However, Table~\ref{table:parameters} shows that the effect of anisotropic part is larger.  More specifically, the magnetiudes of $K_2$, $\Gamma_2$, and $\Gamma_2'$ are larger than that of $\Gamma_1'$ considered in the conventional model and are much larger than that of $J_2$ considered by \citet{yadavKitaevExchangeFieldinduced2016a} (Table~\ref{table:parameters}). Also, \citet{winterChallengesDesignKitaev2016} calculated the anisotropic second-nearest-neighbor exchange interactions, which are in good agreement with our results; however, they considered second-nearest-neighbor exchange interactions as small and neglected them in their proposed model and theoretical analyses. 

\begin{table*}[ht!]
\centering
\caption{(1) The number of fitting parameters ($N_{\text{param}}$) in each model and (2) the root mean square (RMS) error in the total energy per Ru atom [see Eq.~\eqref{eq:RMSE}]. The last column is the product of (1) and (2), i.\,e.\,, the RMS error multiplied by the penalty of each model, which is the number of fitting parameters.}
\begin{tblr}{
    cells = {c},
    hline{1,6} = {-}{0.08em},
    hline{2} = {-}{0.04em}
}
    Model & $(1) N_{\text{param}}$ & (2) RMS error in total energy / Ru (meV) & (2)$\times$(1)\\
    Conventional & 5 & 0.40 & 2.0\\
    Spin-ice & 3 & 0.53 & 1.6\\
    A-$J_1J_2J_3$ & 12 & 0.10 & 1.3\\
    A-$J_1J_2$-I-$J_3$ & 8 & 0.11 & 0.85\\
\end{tblr}
\label{tab:mse}
\end{table*}

We also compared the number of fitting parameters across the different models in Table~\ref{tab:mse}. Furthermore, since the isotropic part of the second-nearest-neighbor exchange interaction and the anisotropic part of the third-nearest-neighbor exchange interaction are relatively small (Table~\ref{table:parameters}), we propose another model in which the isotropic part of the second-nearest-neighbor exchange interaction and the anisotropic part of the third-nearest-neighbor exchange interaction is dropped from the A-$J_1J_2$-I-$J_3$ model; thus, the number of parameters is reduced from 12 to 8.

As shown in Fig.~\ref{fig:fitting}, the newly proposed anisotropic $J_1$-$J_2$ isotropic $J_3$ (A-$J_1J_2$-I-$J_3$) model describes the DFT total energies as well as the A-$J_1J_2J_3$ model, but with fewer parameters. Also, Table~\ref{table:parameters} shows that the values and standard errors of the fitted parameters of the two models are similar. Thus, we suggest the A-$J_1J_2$-I-$J_3$ model as the minimal model for the description of the DFT total energies. By adding only three more parameters to the conventional model, we reduced the root-mean-square (RMS) error in the total energy from 0.40~meV to 0.11~meV, i.\,e.\,, by 3.6 times. The RMS error is defined as
\begin{equation}
\label{eq:RMSE}
    \text{(RMS error)} = \sqrt{\frac{1}{N_\text{conf}}\sum_{i=1}^{N_\text{conf}} \left[E_\text{model}(i) - E_\text{DFT}(i)\right]^2} \, ,
\end{equation}
where $N_\text{conf}$ is the number of configurations and $E_\text{model}(i)$ and $E_\text{DFT}(i)$ are fitted and DFT total energies per Ru ion of a given configuration $i$, respectively. The RMS errors in the total energy of the four models are shown in Table~\ref{tab:mse}.

Moreover, to penalize the increase in the number of fitting parameters in a model, we multiplied the RMS error of each model by the number of fitting parameters, which is shown in the last column of Table~\ref{tab:mse}. Now, it is clear that our proposed models, especially the A-$J_1J_2$-I-$J_3$ model, have an advantage over the conventional or spin-ice model.

In passing, we note that the fact that the error in the total energy calculated from the model based on the $\mathbf{J}_\text{eff}$ picture [Tab.~\ref{tab:mse}] is small indirectly supports the validity of Eq.~\eqref{eq:def_s_eff}.

Lastly, according to our DFT calculations, the ground state is in the zigzag antiferromagnetic configuration with $\theta = 55 \degree$ and $\phi = 0 \degree$ [see Fig.~\ref{fig:fitting}; especially, Figs.~\ref{fig:fitting}(e) and~\ref{fig:fitting}(g)]. This configuration is the same as the experimentally measured ground state configuration~\cite{zigzag_conf,searsFerromagneticKitaevInteraction2020}. Since our model is fitted to this DFT total energy with good accuracy, it naturally reproduces the ground-state configuration [see Figs.~\ref{fig:fitting}(e) and~\ref{fig:fitting}(g)].

% We also note that the computed exchange parameters are from semi-classical approximations, that is, spin operators are considered as classical vectors. In Ref.~\cite{torelliFirstPrinciplesHeisenberg2020}, the authors argue that the DFT total energies correspond to quantum eigenenergies of the quantum spin Hamiltonian. Although Ref.~\cite{torelliFirstPrinciplesHeisenberg2020} discusses only the isotropic case, we computed the quantum ground state energy of quantum spin Hamiltonian from the semi-classical parameters by exact diagonalization.

% The quantum ground state energy is, of course, lower than the energy under the semi-classical approximation. This means that to match the DFT total energy and quantum eigenenergy, the computed exchanges must be modified. Thus, we note that the detailed values of exchanges are subject to changes from quantum effects. 

\subsection{Magnetic moments and $g$ factors}

\begin{table*}[ht]
\centering
\caption{Ideal and computed magnetic moments obtained from atomic orbital projection method}
\begin{tblr}{
  width = 0.7\linewidth,
  colspec = {c c *{3}{X[c]}},
  cell{1}{3} = {c=3}{},
  vline{2-4} = {1}{0.04em},
  vline{2,3} = {1-6}{0.04em},
  hline{1,6} = {-}{0.08em},
  hline{2-3} = {-}{0.04em},
}
 & Ideal $J_{\rm eff}=1/2$ state & DFT(+$U$) + atomic orbital projection \\
Collinear direction $\nu$ & $XYZ$ (isotropic) & $X$ & $Y$ & $Z$  \\
$2S^{[\text{zAFM-}\nu]}_{i,\mu}$ ($\mu_{\rm B}$) & 1/3 & 0.325 & 0.274 & 0.260 \\
$L^{[\text{zAFM-}\nu]}_{i,\mu}$ ($\mu_{\rm B}$) & 2/3 & 0.485 & 0.455 & 0.465 \\
$M^{[\text{zAFM-}\nu]}_{i,\mu}$ ($\mu_{\rm B}$) & 1 & 0.810 & 0.729 & 0.725 \\
% Ratio & 2 & 1.490 & 1.659 & 1.791 & 1.472 & 1.479 & 1.524 & 1.477 & 1.788 & 1.761                             
\end{tblr}
\label{table:moments_atomic}
\end{table*}

\begin{table*}[ht]
\centering
\caption{Computed magnetic moments obtained from moment canting method}
\begin{tblr}{
  cells = {c},
  vline{2} = {-}{0.04em},
  vline{4,6} = {-}{0.04em},
  hline{1,5} = {-}{0.08em},
  hline{2} = {-}{0.04em},
}
\makecell{$1\degree$ canting direction $\mu$ \\(initial direction $\nu$)} & $X\,\,(Y)$ & $X\,\,(Z)$ & $Y\,\,(Z)$ & $Y\,\,(X)$ & $Z\,\,(X)$ & $Z\,\,(Y)$ \\
$\displaystyle \frac{\Delta_{\mu} 2S^{[\text{zAFM-}\nu]}_{\rm tot, \mu}}{N_{\rm Ru}\Delta \theta}$ ($\mu_{\rm B}$) & 0.457 & 0.457 & 0.433 & 0.443 & 0.323 & 0.342  \\
$\displaystyle \frac{\Delta_{\mu} L^{[\text{zAFM-}\nu]}_{\rm tot, \mu}}{N_{\rm Ru}\Delta \theta}$ ($\mu_{\rm B}$) & 0.673 & 0.677 & 0.660 & 0.654 & 0.578 & 0.602  \\
$\displaystyle \frac{\Delta_{\mu} M^{[\text{zAFM-}\nu]}_{\rm tot, \mu}}{N_{\rm Ru}\Delta \theta}$ ($\mu_{\rm B}$) & 1.131 & 1.134 & 1.093 & 1.097 & 0.900 & 0.944 \\
% Ratio & 2 & 1.490 & 1.659 & 1.791 & 1.472 & 1.479 & 1.524 & 1.477 & 1.788 & 1.761                           
\end{tblr}
\label{table:moments_canting}
\end{table*}

We now move on to the computation of magnetic $g$ factors of the monolayer. We present the computed moments. In the ideal $J_\text{eff}=1/2$ picture, $|\vb{S}_i| = 1/6$, $|\vb{L}_i| = 2/3$, and the total moment is $|\vb{L}_i + 2\vb{S}_i|=1$~\cite{jeff_half_state} which results in an isotropic $g=2$ within the ideal $J_\text{eff}=1/2$ picture.
We compared the computational results obtained with two different methods in Tables~\ref{table:moments_atomic} and~\ref{table:moments_canting}. Comparing the two methods shows that the moment-canting method gives about 1.3 times larger moments than the projection method. In other words, the atomic orbital projection misses a non-negligible portion of the total magnetic moment. We also note that the moment-canting method gives similar values for the same canting direction. This supports the validity of the moment-canting method for obtaining the $g$ factor.

%\begin{figure}[ht!]
%    \centering
%    \includegraphics[width=0.45\textwidth]{./figures//figure_distortion.pdf}
%    \caption{The two types of structural distortion that are present in the fully relaxed RuCl$_3$ structure. The red and blue arrows represent the displacement pattern of the Cl atoms (green spheres) in the lower and upper Cl layers, respectively, with respect to the structure consisting of edge-sharing perfect RuCl$_6$ octahedra.}
%    \label{fig:distortion_figure}
%\end{figure}

\begin{table*}[ht]
\centering
\caption{Analysis of the atomic orbital projection method. The N\'eel vector is constrained to be the $X$ direction (Fig.~\ref{fig:crystal_structure}).}
\begin{tblr}{
  cells = {l},
  cell{7}{2} = {c=6}{},
  cell{7}{8} = {c=4}{},
  cell{8}{2} = {c=6}{},
  cell{8}{8} = {c=4}{},
  cell{9}{2} = {c=6}{},
  cell{9}{8} = {c=4}{},
  hline{1,13} = {-}{0.08em},
  hline{2,5,7,10} = {-}{0.04em},
  vline{2,12} = {-}{0.04em},
  vline{8} = {7-9}{0.04em}
}
Hubbard eigenvector & 1 & 2 & 3 & 4 & 5 & 6 & 7 & 8 & 9 & 10 & Total \\
Probability in $J_\text{eff}=3/2$ & 0.173 & 0.964 & 0.865 
& 0.981 & 0.989 & 0.000 & 0.007 & 0.007 & 0.007 & 0.007 & 4.00 \\
Probability in $J_\text{eff}=1/2$ & 0.826 & 0.031 & 0.130 & 0.012 & 0.002 & 1.000 & 0.000 & 0.000 & 0.000 & 0.000 & 2.00 \\
Probability in $e_g$ & 0.001 & 0.005 & 0.005 & 0.007 & 0.009 & 0.000 & 0.993 & 0.993 & 0.993 & 0.993 &  4.00 \\
Electron occupation & 0.99 & 0.99 & 0.99 & 0.98 & 0.98 & 0.31 & 0.63 & 0.63 & 0.62 & 0.62 & 7.72 \\
Hole occupation & 0.01 & 0.01 & 0.01 & 0.02 & 0.02 & 0.68 & 0.37 & 0.37 & 0.38 & 0.38 & 2.28 \\ 
\makecell{Approximate \\ spin moment ($\hbar$)} & \makecell{  0 (Kramer doublets) $+$ [ 0.68 (Hole occupation) \\ $\times$ 1/3 (Spin moment of $J_\text{eff}=1/2$) ] = 0.227} & & & & & & 0 (Kramer doublets) & & & & 0.233 \\
\makecell{Approximate \\ orbital moment ($\hbar$)} & \makecell{ 0 (Kramer doublets) $+$ [ 0.68 (Hole occupation) \\ $\times$ 2/3 (Orbital moment of $J_\text{eff}=1/2$) ] = 0.453} & & & & & & 0 (Kramer doublets) & & & & 0.467\\
\makecell{Approximate \\ total moment ($\hbar$)} & $0.227 + 0.453 = 0.680$ & & & & & & 0 (Kramer doublets) & & & & 0.700 \\
\makecell{Calculated \\ spin moment ($\hbar$)} & & & & & & & & & & & 0.325 \\
\makecell{Calculated \\ orbital moment ($\hbar$)} & & & & & & & & & & & 0.485 \\
\makecell{Calculated \\ total moment ($\hbar$)} & & & & & & & & & & & 0.810
\end{tblr}
\label{table:atomic_projection}
\end{table*}

In order to examine how well the $\jeff${} picture describes the DFT results, we further discuss the atomic orbital projection results by analyzing the eigenvalues and eigenvectors of the Hubbard occupation matrix (Table~\ref{table:atomic_projection}). The occupation matrix has five eigenvalues close to 1, four eigenvalues close to 0.6, and one eigenvalue close to 0.3.
The summation of the electron occupations across the five states with eigenvalues close to 1 gives about four electrons in $J_\text{eff}=3/2$ states and about 1 electron in $J_\text{eff}=1/2$ states. The single state with eigenvalue close to 0.3 is close to a pure $J_\text{eff}=1/2$ state. We now assume a hypothetical situation in which the six mentioned states are all fully occupied. Then, we can mix the states arbitrarily and form three fully occupied Kramer doublets. Due to time-reversal symmetry, these fully occupied doublets do not contribute to spin, orbital, or total magnetic moments. We then return to the real situation; because of the 0.3 occupancy of the single state, we can interpret the six states as three Kramer doublets occupied by holes with occupancy 0.7 in a $J_\text{eff}=1/2$ state. These holes lead to spin magnetic moment $0.7 \times 1/3 = 0.233$ and orbital magnetic moment $0.7 \times 2/3 = 0.467$, similar to the values found by the atomic orbital projection method:  spin = 0.325, orbital = 0.485 (see the last three rows of Table~\ref{table:atomic_projection}). The remaining four states with eigenvalues close to 0.6 are found to be $e_g$ states. These states can also be mixed arbitrarily because of their almost identical occupations and can form two Kramer doublets with occupations of 0.6 for each state. 

\begin{table}[ht!]
\centering
\caption{The calculated $g$ factors of monolayer $\alpha$-RuCl$_3$. Compare with the measured $g$ factors listed in Table~\ref{tab:previous_g}.}
\begin{tblr}{
    cells = {c},
    hline{1,5} = {-}{0.08em},
    hline{2} = {-}{0.04em}
}
    Method & $g_{XY}$ & $g_{Z}$ \\
    Ideal $J_{\rm eff}$ model & 2 & 2 \\
    Atomic orbital projection & 1.54 & 1.45 \\
    $J_{\rm eff}$ moment canting & 2.22 & 1.84
\end{tblr}
\label{tab:our_g}
\end{table}

% TYK comment on g factor
We computed the magnetic $g$ factor from these results. In Table~\ref{tab:our_g}, the $g$ factors obtained from the atomic orbital projection method are $g_{XY} = 1.54$, and $g_Z = 1.45$. The $g$ factors obtained from the moment-canting Wannier interpolation method are $g_{XY}= 2.22$, $g_Z = 1.84$, close to the value if the system was to be described by the ideal $J_{\rm eff}$ model, $g=2$, while the values obtained from the atomic orbital projection method are much smaller. Thus, an accurate computation of the spin and orbital magnetization is needed for an accurate computation of the magnetic $g$ factor. %Also, the anisotropy in the $g$ factors is small, validating the results of Refs.~\onlinecite{majumder_anisotropic_2015,PhysRevB.96.161107,loidl_proximate_2021}. We also note that most of the measured $g$ factors are larger than our values (see Tab.~
%\ref{tab:previous_g}), which calls for further experimental research.

Some earlier studies claimed a large anisotropy between $g_{XY}$ and $g_{Z}$ ($g_{XY}$ = 2.5 and $g_{Z}$ = 0.4 in Ref.~\onlinecite{kubota_successive_2015}  and $g_{XY}=2.51$ and $g_Z=1.09$ in Ref.~\onlinecite{yadavKitaevExchangeFieldinduced2016a}), while the most recent one reported a smaller anisotropy ($g_{XY}$ =  2.9 and $g_{Z}$ = 2.4 in Ref.~\onlinecite{loidl_proximate_2021} (see Table~\ref{tab:previous_g}). Note that the $g$ factor in the ideal $J_\text{eff} = 1/2$ case is expected to be isotropic ($g_{XY}=g_{Z}=2$). This discrepancy between different experiments has existed for many years, as various studies used different models to account for the results.
For instance, the large anisotropy in the $g$ factor as reported by \citet{kubota_successive_2015} is based on the anisotropy in the magnetic susceptibility and high-field magnetization. On the other hand, \citet{searsFerromagneticKitaevInteraction2020} proposed that the large anisotropy in the measured magnetic susceptibility can be explained with large $\Gamma_1$, not resorting to $g$-factor anisotropy. They demonstrated that models with similar magnitudes of $K_1\, (< 0)$ and $\Gamma_1\, (> 0)$ can fit the anisotropic magnetization data, without the need for a large $g$-factor anisotropy. Our calculation results seem to support this smaller anisotropy scenario.

%ne might ask why the value of $g$ factors are almost identical with the ideal $J_\text{eff}=1/2$ while the orbital-spin ratio is different. This is because the total magnetic moment of $e_g$ states is 1, which is the same with the $J_\text{eff}=1/2$ states. That is, the contribution of the $e_g$ state to the $g$ factors is the same with the $J_\text{eff}=1/2$ states, while the orbital-spin ratio is different.

\subsection{Two types of structural distortion}

\begin{figure}[ht!]
    \centering
    \includegraphics[width=0.45\textwidth]{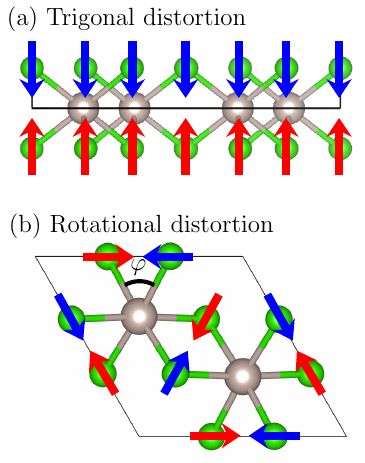}
    \caption{The two types of structural distortion that are present in the fully relaxed RuCl$_3$ structure. The red and blue arrows represent the displacement pattern of the Cl atoms (green spheres) in the lower and upper Cl layers, respectively, with respect to the structure consisting of edge-sharing perfect RuCl$_6$ octahedra.}
    \label{fig:distortion_figure}
\end{figure}

\begin{figure}[ht!]
    \centering
    \includegraphics[width=0.5\textwidth]{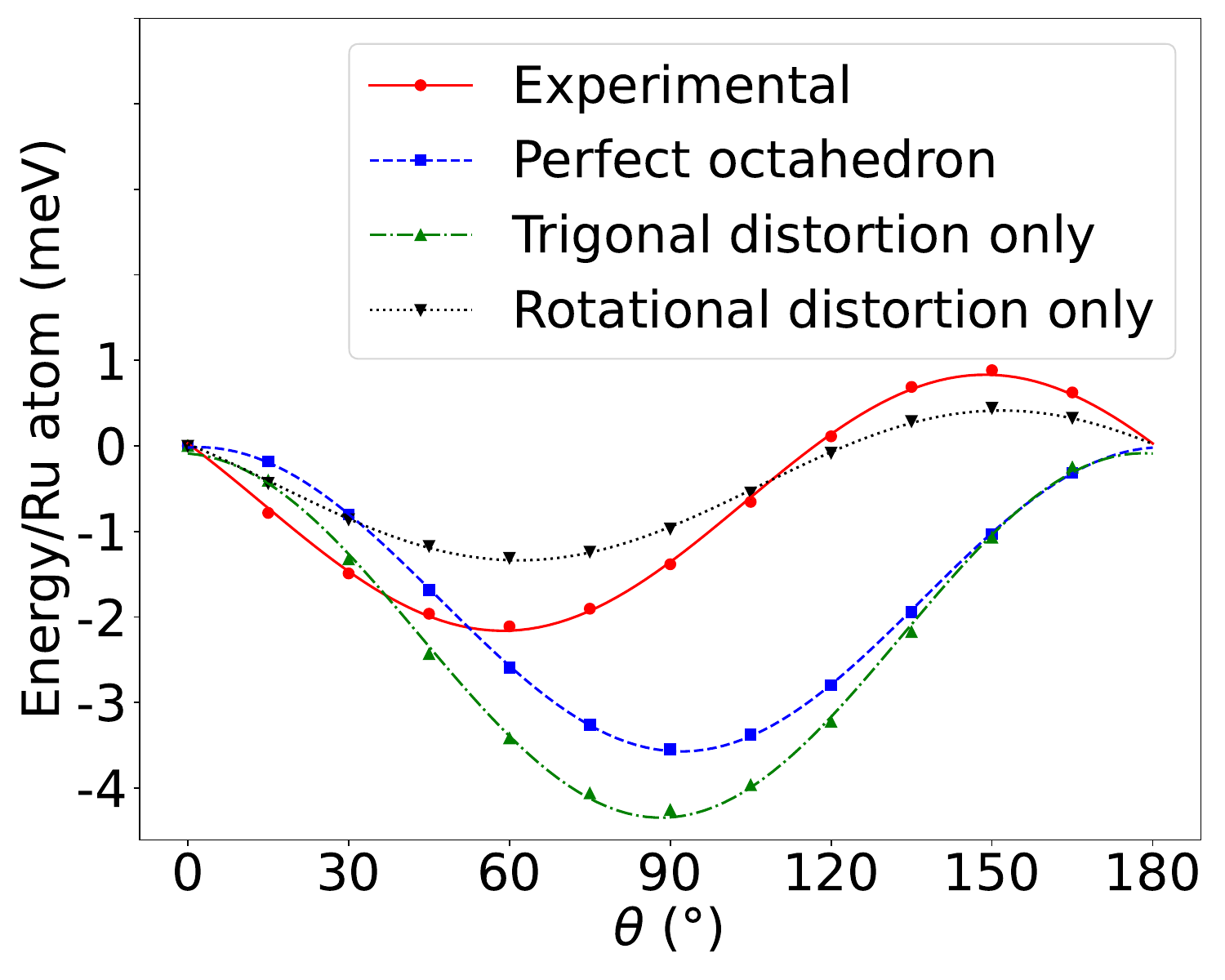}
    \caption{The total energy versus the effective moment direction of monolayer $\alpha$-RuCl$_3$ with the zigzag spin configuration [Fig.~\ref{fig:configurations}(a)] obtained from DFT calculations (symbols) assuming different structures. The energies for the $\theta = 0 \degree$ case were set to 0. The curves are a guide to the eye.} %We used $a + b\sin^2 \theta + c \sin \theta \cos \theta + d \cos^2 \theta$ function for the fitting (curves).}
    \label{fig:distortion}
\end{figure}

Lastly, we discuss the effects of local structural distortion around the RuCl$_6$ complexes on the magnetic properties. Specifically, we considered the trigonal and twist distortions of the Cl atoms with respect to the perfect octahedral coordination (Fig.~\ref{fig:distortion_figure}). For trigonal distortion, the experimental value of the trigonal angle is $55.53 \degree$, which is larger than $\cos^{-1} (1/\sqrt{3}) = 54.74 \degree$ for the perfect octahedron (See Table~\ref{table:relaxation_details}). For twist distortion, the angle $\varphi$ in Fig.~\ref{fig:distortion} is $55.16 \degree$, which is smaller than $60 \degree$ for the perfect octahedron. 

Figure~\ref{fig:distortion} shows one of the main results of this study. We first see that the lowest-energy states are at $\theta\sim60\degree$ and $\theta\sim90\degree$ for the experimental and perfect octahedron structures, respectively, highlighting the importance of using accurate geometries to obtain the correct moment direction ($\theta\sim60\degree$). To examine what type of distortion is mainly responsible for yielding the correct magnetic anisotropy, we repeated the same calculation, imposing either a trigonal or twist distortion only on the perfect octahedron structure. If only trigonal distortion of the chalcogen atoms is present, the angular position of the energy minimum is similar to that of the ideal octahedron case. However, if only the twist distortion of the chalcogen atoms within the $XY$ plane is present, the location of the energy minimum is similar to that of the experimental case. These results demonstrate that the twist distortion of the chalcogen atoms within the $XY$ plane is essential in determining the magnetic properties of $\alpha$-RuCl$_3$.  %We thus suggest that an accurate extraction of the magnetic exchanges from the hopping integrals must include the effect of twist of the chalcogen atoms within the $XY$ plane, which has not been explicitly considered in theoretical models so far.

This finding calls for further theoretical and experimental research. First, previous theoretical studies on extracting the exchange parameters and proposing model Hamiltonians of $\alpha$-RuCl$_3$~\cite{Rau2014_1, Rau2014_2} have considered only the effects of the trigonal distortion but not this twist distortion that we found.  We expect that the incorporation of the twist distortion at the level of the model electronic Hamiltonian would lead to a more accurate theory for the honeycomb magnetic systems. Second, such theoretical models that incorporates the twist distortions can provide further insight into the interpretation of high-pressure experiments~\cite{high_pressure_1, high_pressure_2} and future experiments on $\alpha$-RuCl$_3$.
%This also suggests the possible reason for the discrepancy of our orbital-spin ratio and Ref.~\cite{PhysRevB.96.161107}. In the latter, the distortions beyond the trigonal one are neglected in the multiplet calculations.

\section{Conclusion}
We demonstrated that the low-temperature ground state stacking of bulk RuCl$_3$ is of $R\bar{3}$ type. We then showed that the electronic states near the band gap are well described by the $J_\textrm{eff}$ states by choosing the angular momentum quantization axis to be parallel to the N\'eel vector. We also constructed a comprehensive effective moment model for monolayer $\alpha$-RuCl$_3$ that takes into account Kitaev-type anisotropy and more, and obtained the anisotropic $g$ factor from first-principles calculations. We applied a constrained DFT formalism to Hubbard projected $\jeff${} moments and obtained the exchange parameters using the energy mapping method. In addition, we computed the magnetic $g$ factors from the $\jeff${} moment canting method. We showed that anisotropic second-nearest-neighbor exchange terms were crucial in reproducing the total energies from the constrained DFT calculations. Furthermore, we applied the translationally equivariant Wannier interpolation of orbital magnetization to $\alpha$-RuCl$_3$ and showed that the atomic orbital projection method was inappropriate for accurate computation of the magnetic $g$ factors. Finally, we examined the effect of structure on the magnetic properties and concluded that considering the so far neglected twist of the chalcogen atoms within the $XY$ plane was essential for the correct description of the magnetism of $\alpha$-RuCl$_3$. Our work thus proposes a different viewpoint on the magnetism of $\alpha$-RuCl$_3$, and suggests further directions for research on this material.

\begin{acknowledgments}
We thank Kwang-Yong Choi for helpful discussions. This work was supported by the Korean NRF No-2023R1A2C1007297 and the Institute for Basic Science (No. IBSR009-D1). Computational resources were provided by KISTI (KSC-2023-CRE-0533).
\end{acknowledgments}

\appendix
\counterwithin*{figure}{section}
\renewcommand\thefigure{\thesection\arabic{figure}}
\counterwithin*{table}{section}
\renewcommand\thetable{\thesection\Roman{table}}

\bibliography{main}

\end{document}